# High Accuracy Protein Identification: Fusion of solid-state nanopore sensing and machine learning

Shankar Dutt[1, γ, *], Hancheng Shao[1, γ], Buddini Karawdeniya[2], Y. M. Nuwan D. Y. Bandara[3], Elena Daskalaki[4], Hanna Suominen[4,5], Patrick Kluth[1]

[1]Department of Materials Physics, Research School of Physics, Australian National University, Canberra ACT 2601, Australia

[2]Department of Electronic Materials Engineering, Research School of Physics, Australian National University, Canberra ACT 2601, Australia

[3]Research School of Chemistry, Australian National University, Canberra ACT 2601, Australia

[4]School of Computing, College of Engineering, Computing and Cybernetics, Australian National University, Canberra ACT 2601, Australia

[5]Eccles Institute of Neuroscience, College of Health and Medicine, Australian National University, Canberra ACT 2601, Australia

γ: These authors contributed equally.

*Email: shankar.dutt@anu.edu.au

# ABSTRACT

Proteins are arguably one of the most important class of biomarkers for health diagnostic purposes. Label-free solid-state nanopore sensing is a versatile technique for sensing and analysing biomolecules such as proteins at single-molecule level. While molecular-level information on size, shape, and charge of proteins can be assessed by nanopores, the identification of proteins with comparable sizes remains a challenge. Here, we combine solid-state nanopore sensing with machine learning to address this challenge. We assess the translocations of four similarly sized proteins using amplifiers with bandwidths (BWs) of 100 kHz and 10 MHz, the highest bandwidth reported for protein sensing, using nanopores fabricated in <10 nm thick silicon nitride membranes.  F-values of up to 65.9% and 83.2% (without clustering of the protein signals) were achieved with 100 kHz and 10 MHz BW measurements, respectively, for identification of the four proteins. The accuracy of protein identification was further enhanced by classifying the signals into different clusters based on signal attributes, with F-value and specificity of up to 88.7% and 96.4% respectively for combinations of four proteins. The combined use of high bandwidth instruments, advanced clustering, machine learning, and other advanced data analysis methods allows label-free identification of proteins with high accuracy.

# 1. INTRODUCTION

Proteins are the vital building blocks of life, orchestrating a vast array of biological functions and processes that maintain health and critical cellular functions. They can serve as biomarkers for diagnosis and monitoring of diseases, facilitate cellular signaling and reaction catalysis, and enable transport and storage of critical ions and molecules[1–6]. Proteins are constructed of amino acid chains folded into specific tertiary and quaternary structures that govern their function. While proteins are vital actors for biological processes and functions essential for life, their presence or fluctuations of their typical levels can also indicate detrimental biological processes such as adverse health conditions. In some circumstances, proteins contribute to the advancement of diseases by intensifying their activity to create favorable settings that enhance the progression of the disease, e.g. proteins such as PADI4 and HIF-1 facilitate the growth of cancer[7–9]. Irrespective of the function, selective protein detection and quantification is critical for the identification, evaluation and understanding of biological processes and related progress, e.g., for future drug development.

Complex biological samples like serum, saliva, and urine contain a multitude of protein biomarkers indicative of a host of health conditions. There are numerous conventional analytical methods for protein detection, characterization, and quantification such as mass spectrometry (MS)[10,11], protein NMR spectroscopy[12], enzyme-linked immunosorbent assay (ELISA)[13], protein immunoprecipitation[14], X-ray crystallography[15], and fluorescence resonance energy transfer (FRET)[16]. While these have seen widespread adoption as tools for protein profiling, inherent challenges include tedious sample preparation and complex instrumentation (MS), protein labeling (FRET), the need for specific receptors (ELISA), biologically active crystalline states (X-ray), and lack of sufficient sensitivity for some techniques. A label-free, portable technology with minimal sample preparation, with the ability to operate in biomimetic fluids for selective detection of low-abundance targets (i.e., for early detection) in complex samples could be transformative (i.e., many tests in one platform). While lateral-flow-assay-based point-of-care devices tick most of these boxes, they are designed for the visual detection of a specific target using specific receptors and may not be ideal for the early detection of a health condition (i.e., low abundance assaying).

In this study, we used a solid-state nanopore (nanopore hereafter) sensor and machine learning (ML) for the selective identification of four similar-sized proteins. A nanopore sensor, in its simplest definition, is a nanoscale aperture spanning an otherwise impervious membrane separating two electrolyte reservoirs. The analyte is added to one side (*cis* side) and a suitable voltage is applied to the other side (*trans* side) to drive the molecules across the nanopore generating analyte-specific information as resistive or conductive pulses. Importantly, nanopore sensing can be conducted under various electrolyte chemistries and the sample is typically added as-is. The versatility of this tag-free technology is well demonstrated by its application repertoire spanning a host of biological classes such as DNA[17,18], proteins[19,20], glycans[21,22], viruses[23,24], and liposomes[25]. Unsurprisingly, nanopores have also demonstrated great potential in studying the fundamental protein structure as well as its biochemistry and biophysics; e.g., protein flexibility[26], folding-unfolding[27], molecular weight[28], conformational differences[29], and interaction with proteins[30] have been studied using nanopores. A key challenge in protein sensing is the fast translocation speed associated with proteins. With the more ubiquitous Axopatch 200B amplifier in the resistive-feedback mode, translocations with residence times less than 10 μs are attenuated[28]. Thus, there is a constant interest to perform nanopore sensing experiments using MHz-level bandwidth (BW) equipment as they allow sub-microsecond resolution readouts and, in this study, we report the detection of proteins with bandwidths up to 10 MHz (sampling rate of 40 Msps). The caveat with higher bandwidth is the increase in open-pore current noise which can be circumvented partially by noise reduction (e.g., membranes fabricated employing low dielectric noise materials such as quartz[31]) and signal magnitude enhancement (e.g., thinner membranes) strategies. For nanopore-based protein sensing experiments, the highest bandwidth reported to date is 1 MHz[32], and for DNA bandwidths up to 10 MHz[31,33] has been reported. The high bandwidth is instrumental for the distinction between similar sized proteins and the complexity of the problem calls for using ML for a more advanced data analysis.

One of the main challenges of nanopore sensing has been the selective identification of analytes in a complex mixture by electrical readouts alone, i.e., uniquely associating the resistive/conductive pulse of the nanopore sensor to a specific analyte. While modifying the nanopore surface with a selective receptor such as an antibody or an aptamer could incorporate the much-needed selectivity, the shelf-life of the functional layers, device-to-device variability, sensitivity/susceptibility to electrolyte conditions, biofouling (in the case of complex-samples) and limited throughput could diminish the anticipated selectivity. The lack of selectivity is exacerbated by the fact, despite being a single molecule sensor, the individual current drop signals resulting from translocations have so far been mainly characterized by two parameters: the pulse width (i.e., translocation time; $\Delta t$) and pulse depth ($\Delta I$). In certain instances, the area of the event, noise levels and stepwise information of events has also been used for analysis which are correlated to the pulse width and pulse depth[34,35]. While these parameters can generate a fingerprint of a molecule leading to positive identification, for molecules of very similar weight, size, and charge in a complex mixture, accurate identification with such simple metrics alone is hardly possible. This type of conventional analysis leads to an enormous loss of data since each signal is just characterized by two features (i.e., $\Delta t$ and $\Delta I$). Using an increasing number of features of the signal or even the entire signal renders conventional statistical analysis extremely difficult which is further exasperated by the number of different signals that can result from one analyte.

Thus, supervised ML approaches have become attractive for analyte identification. The use of such ML approaches has gained tremendous traction in DNA sequencing[36], and the determination of structural composition of polysaccharides[37]. Additionally, identification of larger particles such as bacteria and viruses are also emerging triggered by the current pandemic and other global disease outbreaks. This is exemplified by the works of Tsutsui *et al.* [38] (bacteria) and, Taniguchi *et al.* [23] (viruses) where solid state nanopores and ML methods have been combined to discriminate between single-bacterial shapes and to identify coronaviruses, respectively. For proteins, Raynaud *et al.* [39] have used seven features to distinguish a binary protein mixture, currently the only existing ML-incorporated protein sensing study using functionalized nanopores. Here we present a quaternary protein system with data acquisition in the 10 MHz BW domain (compared to the conventional 100 kHz) and >29,000 events on average (for different proteins measured at different applied bias) (~76 × higher than previously mentioned study). Nanopore data undoubtably deliver sufficient information to differentiate proteins (biomolecules in general), yet as evident by the literature, classical analysis techniques are not deciphering the additional layers of information encoded in nanopore signals. In this study, we explore ML coupled with solid-state nanopores for identification and discrimination of proteins to elevate the potential of this already powerful single-molecule sensing tool for protein sensing.

## 2. RESULTS

### 2.1 Fusion of nanopore sensing and machine learning

Figure 1(a) shows a schematic of protein translocation through a ~7 nm thick silicon nitride membrane. The translocation measurements were done using two portable amplifiers developed by Elements SRL (Figure 1 (b)) with bandwidths of 100 kHz (200 ksps) and 10 MHz (40 Msps). In this investigation, four proteins (Hb, HSA, BSA, and Con A) were considered because of their similar molecular weights and/or sizes (Table S1). Proteins with significantly different sizes and/or molecular weights will generate distinct electrical readouts and are thus relatively easy to distinguish. The choice of proteins was therefore made to imitate a challenge that arises when trying to identify similar-sized proteins in complex real-world solutions, such as blood serum. Hb, HSA and BSA have comparable molecular weights. The resemblance between HSA and BSA extends further than just their molecular weights; they also share 76% sequence identity[40]. Furthermore, both HSA and BSA exhibit a prolate shape, with nominal sizes at pH ~ 8 of 7.1 nm and 7.2 nm, respectively. Con A on the other hand, manifests as a spherical homotetramer, with a molecular weight of 102 kDa. The size of Con A, is approximately 8 nm in diameter at pH ~8, which is close to that of BSA and HSA[29]. Representative current-time traces from translocation measurements at BW of 100 kHz and 10 MHz are shown in Figure 1(d) and (e) respectively.

Unlike the other three proteins, measurement of Hb under higher applied voltages often leads to protracted/irreversible pore occlusions that either require user intervention or solution exchange to recover the open-pore status. This made the collection of a statistically significant data set without a change in pore characteristics due to adsorption of proteins at 600 mV challenging. After repeated attempts, we limited the applied voltage for the Hb measurements to 500 mV unlike other proteins studied in this work. Figure 1 (f) and (g) illustrate an example of unfiltered signals produced by measurements at 100 kHz and 10 MHz BW, respectively followed by lowpass filtering with various cut-off frequencies. As anticipated, lowering the cut-off frequency lowers the measurement's total noise, but at the expense of signal quality necessary for ML-based classification. The intricacies of the signal structure are almost completely lost at 10 kHz cut-off frequency emphasizing the need for higher bandwidth instrumentation for nanopore sensing[20,41,42]. While this frequency has its merits for conventional two-metric analysis based on Δt and ΔI, the loss of intricate intra-signal details makes it inefficient for the current study. For measurements performed at BW of 100 kHz and 10 MHz, we thus chose cut-off frequencies of 35 kHz and 100 kHz, respectively, because these frequencies had the greatest signal-to-noise ratios while still maintaining the fine features of the signals.

Under an applied bias of 500 mV, the histograms for the drop in ionic current as proteins translocates through the nanopore ($\Delta I$), dwell time ($\Delta t$), and scatter plots exhibiting $\Delta I$ as a function of $\Delta t$ are shown in Figures 2(a) and (e) for all proteins. Other voltage-specific histograms and scatter plots are provided in the SI (figures S1, S2, S4, S5). The columnar pattern discernible in the scatter plots, in particular evident in the measurements conducted at a 100 kHz bandwidth, is due to the signal digitisation process. As can be observed in the scatter plots, the collected signal with different bandwidths under otherwise similar conditions show distribution of dwell times noticeably different from each other. This is expected considering that the Elements 10 MHz amplifier can operate at a 100× higher bandwidth and 200× the sampling rate compared to the Elements 100 kHz amplifier. For ease of comparison, histograms corresponding to $\Delta I$ for the four proteins investigated with the 100 kHz instrument (200 ksps, BW=100 kHz, LPF=35 kHz) and the 10 MHz instrument (40 Msps, BW=10 MHz, LPF=100 kHz) are shown in Figures 2(b)-(d) and (f)-(h). We observe significant overlap of the histograms because the protein sizes or molecular weights are similar. A closer look reveals that the histograms change with applied bias, and under certain conditions, two distributions are visible, which highlights the dependence of protein conformations on the applied voltage. Figures S3 and S6 depict the overlapping histograms corresponding to the change in dwell time for different applied biases, further supporting the usage of ML for this challenge.

ML was performed to identify the proteins using the methods described in the Methods section (Figure 3a). As expected, the quality of the input(s) to the algorithm determines how well the classifier performs. Five distinct feature extraction procedures were created for each of the schemes (figure 3b, table 1) to extract the features in segmented and whole signals needed for ML. The confusion matrices produced by the machine learning classification corresponding to the four proteins by measurements taken at 100 kHz and 10 MHz BWs in response to 500 mV are shown in Figure 3(c) and (d). Figure 3 (e) and (f) demonstrate the corresponding F-values for dual, tri and quaternary protein systems. For this data, feature extraction scheme 3 (see table 1) was applied. The fine features of signals are more apparent with the 10 MHz amplifier compared to its 100 kHz counterpart due to significant bandwidth difference and the ability of the former to sample data at 40 Msps while the latter is restricted to 200 ksps: a 50 µs long event would be portrayed by 2000 points with the 10 MHz amplifier while with the latter it would be limited to 10 data points. These differences are well reflected in the F-values where the result obtained with at 10 MHz bandwidth instrument was 78.9 ± 0.2%, as opposed to 65.9 ± 0.3% with the 100 kHz bandwidth instrument. A similar difference is observed in the sensitivity values, where we obtain 79.4 ± 0.3% sensitivity from measurements done at 10 MHz bandwidth as compared to 65.3 ± 0.3% from measurements done at 100 kHz BW. Additionally, as expected, noticeable differences are also observed in the precision values (78.8 ± 0.2% for 10 MHz BW measurements compared to 66.7 ± 0.3% for 100 kHz BW measurements) as well as specificity values (93.1 ± 0.3 % for 10 MHz BW measurements and 88.4 ± 0.3% for 100 kHz BW measurements). The impact of measurements under different bias on the F-values is discussed below. For measurements done at 500 mV, our results uncovered a noteworthy decrease in the true positive rate associated with HSA when measuring at a bandwidth of 10 MHz as compared to 100 kHz. This result can be attributed primarily to the substantial presence of a secondary population that seems to overlap with populations of the other proteins observed within the data obtained from the 10 MHz bandwidth (see figure 2 (h)). Notably, this overlapping secondary population, and consequently the influence it imparts, is markedly reduced when conducting measurements at the 100 kHz bandwidth. Though not explicitly evident, this disparity may potentially be ascribed to secondary population not been detected due to fast translocations at 100 kHz bandwidth.

## 2.2 Effect of Clustering on Protein Identification

As proteins are complex charged molecules (with both charged and hydrophobic moieties), they often show physiochemical interactions with the nanopore membranes[13]. These interactions (both specific and non-specific) include, for example, the adsorption to the nanopore surface and, protein-protein interactions. Such interactions can be protein specific or can be inherent to the nanopore system. While it is considerably challenging to precisely attribute each signal to its corresponding interaction, it is feasible to correlate certain distinctive signal characteristics to specific types of interaction. For example, in the event of a protein interacting with the nanopore wall, the resulting signal is often deep and prolonged compared to smooth translocations[43]. Influencing factors such as nanopore dimensions, the nature of the biomolecule involved, and the surface charge on the nanopore walls, can cause these events to persist for a duration spanning milliseconds to seconds. Conversely, when a protein undergoes tumbling within the nanopore, it often gives rise to multi-level events characterized by pronounced fluctuations in the intra-event signal[44,45]. Furthermore, if the protein undergoes partial unfolding, induced either by the electric field or physical confinement, the resulting events are typically shallow and protracted. We divided the signals into various individual clusters based on the features of the signals which may be an indication of smooth translocations and different interactions as discussed earlier. These are shown in Figure 4(a) and (b), which correspond to measurements done at BW of 100 kHz and 10 MHz, respectively. The clusters shown correspond to data measured under a bias of 500 mV. Clusters for the other applied biases are given in Figures S7 - S10. In addition to the cluster centres, the distinct signals are also displayed using thin lines with a transparency of 80%. Cluster 0 with the highest population was associated with smooth translocations of the proteins. Although it is difficult to determine which interaction leads to a particular cluster based on this data, proteins moving closer to the pore surface (more pore-surface interactions and by extension higher confinement) should have longer dwell times compared to the translocations along the principal axis (i.e., smooth translocations). For certain clusters, we also see a brief rising time followed by a prolonged crest, which is another sign of a transient change brought on by the physical confinement of the biomolecule or induced by the electric field of the nanopore. We extracted features from the signals associated with specific clusters using Scheme 3 and utilised those features to train and test the data in order to investigate the impact of generated clusters on the identification of proteins. Figure 4(c) shows the F-values for the case of four proteins as a function of signals belonging to different clusters. The blue and red hues represent the F-values from measurements done at 100 kHz and 10 MHz, respectively. It is apparent that cluster 0 has the lowest F-value that we believe to be from smooth translocations. The signals belonging to cluster 1-3 are believed to be the result of different protein-pore interactions and have much higher F-values. In particular, clusters 1 and 2 show F-values of $75.1 \pm 0.4$ % and $73.2 \pm 0.5$%, respectively, for measurements done at 200 ksps. For measurements carried out at 40 Msps, the F-values increased to $85.3 \pm 0.3$% (cluster 1) and $88.7 \pm 0.3$% (cluster 2), demonstrating that the various proteins interact with the nanopore system in very different ways and that the signals produced by these interactions enable more precise discrimination between these four proteins. The highest values for sensitivity and specificity for are also observed with clustering the signals, i.e., $74.9 \pm 0.4$% and $91.6 \pm 0.3$%, respectively, for measurements at 100 kHz BW corresponding to cluster 1 and $88.8 \pm 0.3$% and $96.4 \pm 0.3$%, respectively, for measurements at 10 MHz BW corresponding to cluster 2.

## 2.3 Effect of Feature Extraction Scheme on Protein Identification

As was previously mentioned, the accuracy of the ML prediction depends on the quality of the data fed into the process. As a result, we evaluated the accuracy for the identification of the quaternary-protein system using various feature extraction techniques. The F-values corresponding to the combination of four proteins are displayed in Figure 4(d) as a function of the various schemes. Scheme 3 provided the highest accuracy for measurements done at both 100 kHz and 10 MHz. As compared to Scheme 1, we saw a significant increase in the F-values obtained from Scheme 3 showing the importance of using a multi-feature approach and the significance of features such as area under the curve, the tailedness and the asymmetry of the signal. Scheme 4 was not used with the data resulting from measurements done at 100 kHz BW as at this sampling rate as the signal does not have enough data points to enable the division of the signal into 50 parts. Despite being highly computationally costly, Scheme 5, which uses the entire signal as an input along with some characteristics about the signal shape, did not show increased F-values as compared to Scheme 3 and was thus not employed for the data obtained from measurements carried out at 10 MHz BW.

## 2.4 Effect of Low Pass Filter (LPF) on Protein Identification

Figure 4 (e) shows the impact of different cut-off frequencies of the LPF on the F-values. A 10 kHz LPF is often used for filtering raw data to increase the SNR with some studies using 5 kHz cut-off frequencies[18,46]. However, as shown in Figure 1(f), at these cut-off frequencies (i.e., 5 kHz and 10 kHz), essential information in the signals is lost. The accuracy of the results produced by the classifier is directly impacted by the loss of information as a result of choosing low cut-off frequencies. On the other hand, noise in the signal can also reduce the accuracy if the signal is not adequately filtered. The values shown in Figure 4(e) clearly demonstrate this where an optimal cut-off frequency is needed to maximize the F-values. Deducing the best LPF for a given BW involves striking a balance between noise and signal distortion arising from filtering the signal. While lower LPF settings would decrease noise, it also contributes to signal distortion and thereby leading to loss of characteristic signal features. On the other hand, increasing the LPF cut-off frequency would lead to increase in noise while preserving more of signal characteristics. Since lower noise and preservation of signal characteristics lead to higher F-values, sub-optimal LPF values can lead to lower F-values either due to signal distortion (low LPF) or high noise (high LPF). Thus, a fine tug-of-war between these two parameters defines the best LPF for a given BW (and sampling rate). We observe that the measurements at 10 MHz and 100 kHz BW demonstrated optimal outcomes when filtered at 100 kHz and 35 kHz respectively. For measurements taken at 100 kHz and 10 MHz, the data was thus filtered at 35 kHz and 100 kHz, respectively, and was used to produce all other results described in this study.

## 2.5 Effect of Applied Bias on Protein Identification

We also investigated how the applied voltage during the measurements affected the accuracy of protein identification. The applied voltage has been shown to influence the translocating conformation of proteins. Furthermore, the applied voltage and residence time are inversely correlated with few exceptions such as in instances where voltage-mediated protein unfolding is taking place. To study the effect of voltage on the protein identification, we computed the F-values resulting from measurements performed under different applied biases (i.e., 300mV, 400mV and 500mV). For both the measurements done at 100 kHz and 10 MHz BW, the lowest F-values were obtained for experiments under 400 mV. The highest F-values obtained for 100 kHz and 10 MHz data were $65.9 \pm 0.3$ % (500 mV) and $83.2 \pm 0.4$ % (300 mV), respectively. While one would generally expect to see a higher F-value at lower voltages due to slow translocation speeds, it is interesting to note the lack of any linear correlation with the applied voltage and the F-value in the two cases. Since proteins are sensitive to the voltage bias used for translocation experiments unlike more rigid structures like DNA, this result further emphasizes the need to probe the translocations over multiple voltages rather than choosing an arbitrary voltage bias.

## 3. CONCLUSIONS

Machine learning is highly promising for the identification of similar-sized proteins with high accuracy using single molecule nanopore measurements. Since the measurement platform makes use of solid state nanopores, there is a small tolerance for variation in the diameters of the nanopores between membranes. We were able to translate the trained data from one nanopore and test it with a brand-new set of data from another nanopore and obtained almost identical accuracies (within 3%) owing to the standardisation procedures (using the open pore conductance to normalise the conductance drop) utilised in the current work. The portable 10 MHz bandwidth amplifier yields detailed signals that improve protein identification significantly compared to commonly used 100 kHz amplifiers. We demonstrated that the protein identification accuracy varies with the applied trans-membrane bias which could be associated to the presence of different protein conformations at different voltages. With F-values as high as 99.3%, we observed extraordinarily strong discrimination between proteins of comparable sizes in two protein combinations. With the help of clustering and high bandwidth measurements, F-values and specificity values as high as 88.7% and 96.4% respectively are obtained for combinations of four proteins. The fusion of solid state nanopore sensing and ML is thus very promising for the identification of proteins in complex samples. Selectivity has been a major challenge in label free nanopore sensing and this is an important step towards addressing this challenge. For our study, widely used thin $SiN_x$ membranes were employed. Methodologies to slowdown the translocation of the proteins through solid state nanopores and reduce dielectric noise can yield further improvements in data quality that can further increase accuracy. Our methodology offers the possibility to study variations in post-translational modifications of proteins, and protein-protein interactions that may also provide important insights into the underlying processes of diseases. Further enhancements, such as choice of electrolyte, electrolyte concentration, and employing asymmetric electrolyte concentration in the cis and trans chambers, may further increase the capacity generate more accurate identifications of the proteins.

# 4. METHODS

**4.1 Nanopore fabrication:** Free standing, ~7 nm thick silicon nitride membranes with a ~100 nm thick silicon dioxide underlayer of size 40 μm x 40 μm on a 300 μm thick silicon frame were fabricated as discussed previously[47]. The membranes were placed between two custom built PMMA half cells, with reservoirs containing a 1M KCl electrolyte solution buffered to a pH of ~7. A Keithely 2450 sourcemeter was used to apply an electric field <1 V/nm across the membrane, which was stopped as soon as a rapid surge in current was observed, indicating the creation of a nanopore[48,49]. To estimate the size of fabricated nanopore, a current-voltage (I-V) curve was obtained using the eNPR-200 amplifier: the slope of the curve (i.e., open-pore conductance, G) was then used to estimate the diameter of the pore using,

$$G_0 = K \left( \frac{1}{\frac{\pi r_0^2}{L_0} + \frac{\mu|\sigma|}{K}\frac{2\pi r_0}{L_0}} + \frac{2}{\alpha 2 r_0 + \beta \frac{\mu|\sigma|}{K}} \right)^{-1}$$

Where L and $r_0$ are membrane thickness and pore radius, respectively. The diameters of the nanopores used in this investigation ranged from ~15 to 17 nm (cf. table S2).

**4.2 Analytes and Data Acquisition:** Information about the proteins (Bovine Hemoglobin (Hb), Human Serum Albumin (HSA), Bovine Serum Albumin (BSA), and Concanavalin A (Con A)) used in this study is given in table S1. Figure 1(c) shows the structure of the proteins obtained from the Protein Data Bank website of the Research Collaboratory for Structural Bioinformatics. 1M KCl (Sigma Aldrich, P9333) buffered with 10 mM tris-EDTA were used for all translocation experiments. The target protein was added to the cis side to a final concentration of ~23 nM. Using concentrated drops of HCl (Ajax-Finechem, AJA1367, 36%) or KOH (Chem Supply, PA161), the pH of the electrolyte was adjusted to the desired level and measured using an Orion Star™ pH meter. Figure 1(a) shows the schematic of protein translocation through a nanopore in response to a voltage bias applied to the *trans* side. We used two different portable amplifiers (Figure 1(c)): Elements nanopore readers with maximum bandwidths of a) 100 kHz and b) 10 MHz providing sampling rates of 200 ksps and 40 Msps respectively (these would hereafter be referred to as 100 kHz and 10 MHz amplifiers, respectively). While the 100 kHz amplifier produces data at bandwidths similar to conventional amplifiers such as Axopatch 200B used for the majority of nanopore experiments, the 10 MHz amplifier offers the highest-bandwidth measurements with a temporal resolution of 25 nanoseconds, generating data at a maximum rate of ~8.8 GB/min[31]. As prolate shaped proteins are known to have preferred orientations at different electric fields[29], the measurements were performed at voltages ranging from 300 mV to 600 mV for measurements done at 100 kHz BW and from 300 mV to 500 mV for measurements done at 10 MHz BW. Measurements were not taken at 600 mV using 10 MHz BW amplifier as ionic current corresponding to an applied bias of 600 mV measurements were very close to the current measurement limit (± 100 nA) of the Nanopore Reader 10 MHz.

**4.3 Lowpass filtration and signal extraction:** The acquired data was filtered using a 35 kHz and 100 kHz Butterworth filter for measurements done at 100 kHz BW (200 ksps) and 10 MHz BW (40 Msps), respectively. It was discovered that these cut-off frequencies offered the best signal-to-noise ratio without sacrificing details in the signal (see Results and Discussion section). The event extraction was performed using a custom python-based code using vectorized operations from *numpy* and *bottleneck* libraries (see GitHub repository for the code). An adaptive threshold at or above $5 \times I_{std}$ where $I_{std}$ is the standard deviation of the baseline in the analysis window, was used to flag the pulses to be extracted and stored (i.e., events or signals).

**4.4 Clustering and feature extraction:** The classification of the extracted signals into different clusters (based on a similarity threshold of >85% determined by Pearson correlation) was carried out using a K-means algorithm[50]. Each signal was first assigned to a cluster at random by the K-means algorithm, which then repeatedly improved the clusters by shifting the signals to the cluster whose cluster centre is the most similar to the signal. The cluster centre shifts with each assignment, and the procedure was continued until either the cluster assignments cease shifting or a certain number of iterations had been reached. After different iterations in number of clusters, four clusters were found to be the optimal number necessary for each cluster centre to offer details about various biomolecule translocation conformations through the nanopore without being too similar to other clusters (cf. Figure 4 (a, b)).

To employ both clustered and non-clustered signals for supervised learning, signal characteristics other than conventionally used ΔI and Δt need to be extracted. These features were passed on to the classifier as input. As the shape, size, and surface charge of biomolecules vary not only from one species to another but also within the same species, the extraction of distinguishing features has a significant impact on the accuracy of the results. Figure 3(b) shows a schematic of several resistive pulse feature parameters. In brevity, the resistive pulse is divided into 'n' equally spaced segments along its width where the current drop at the median of each segment ($\Delta i_1$, $\Delta i_2$, … $\Delta i_n$), the pulse width at full width half maximum ($t_{fwhm}$), the pulse width of each segment ($t_0$ /n), the drop in current at full width half maximum ($i_{fwhm}$), the maximum current drop ($i_{max}$), the area under the curve, the kurtosis (measure of asymmetry of the signal), and the skewness (measure of the tailedness of the signal), are used as key features for machine learning. All or some of these features were used for machine learning and the collection of the features used for the machine learning is referred to as a scheme (see Table 1 for more details) We used five different schemes as shown in Table 1 and compared the results obtained from each (discussed in the Results Section).

**4.5 Machine learning:** We implemented supervised ML using random forest[51] and rotation forest[52] classifiers, supported by the large number of input signals (see table S3). These classifiers adopt bagging or bootstrap sampling and scale well with high numbers of uncorrelated trees. Additionally, prior ML investigations involving the detection of viruses using solid-state nanopores have produced promising results using similar approaches[23,24,53]. For the identification of the proteins considered in this investigation, we found that the random forest classifier performed slightly better than rotation forest (with F-values up to 6% higher with the random forest classifier). Thus, results corresponding to the random forest implementations are shown and discussed in this work hereafter. The classifier categorizes the results into four types: true positive, true negative, false positive and false negative and the ratio of these values (see Fig S11) gives the precision, recall/sensitivity, specificity, and F-value (also known as F1-values). In context of this research, 'precision' is defined as the ratio of accurately predicted positive protein identifications to the aggregate of predicted positive protein identifications. The significance of 'precision' lies in its role as an indicator of false detection rates, with higher precision values signifying fewer false positives. Sensitivity, alternatively known as the true positive rate, is a measure of our model's ability to correctly identify proteins. A high sensitivity score indicates that the model excels at detecting all proteins that exist within the solution. On the other hand, specificity is expressed as the ratio of accurately predicted negative protein identifications to the total number of actual negative protein identifications. A model demonstrating high specificity is adept at distinguishing proteins that do not exist in the solution. Within the context of this study, 'positive protein identifications' are those that the model is specifically designed to recognize, while 'negative protein identifications' refer to proteins which the model is constructed to identify as non-target proteins.

For training the algorithm, the data features from different proteins were binarized. The testing sets included a random combination of the data from different experiments of translocation of various proteins through different nanopores. Under-sampling was utilised to balance out the unequal datasets since the number of retrieved signals varies between measurements. This was done by maintaining all data in the minority class and reducing the size of the majority class. Random or sequential selection can be used to choose the signals from the majority class (both options are possible through the provided code). The ML models and their evaluations were implemented in Scikit-learn[54]. An m-fold cross validation method (m = 10) was employed to compute values of the aforementioned evaluation measures. As a result, each domain's data was first divided into training data and test data. The training data was then divided into 10 separate subsets at random. Each subset was used as a validation set once while the other nine subsets were used as training set. Grid Search was implemented to find the best combination of hyperparameters. Classes were balanced during evaluation. The best model from the 10 training attempts was used as final model which was then evaluated on the initial test data. The final F-scores shown in the study were obtained as averages over all regressions and standard deviations from results of different regressions are also reported. For implementing ML on data resulting from different clusters, the raw signals originating from each protein (before the clustering procedure) were randomly divided into two parts with 80:20 split corresponding to raw training data (80%) and test data (20%). At this point, the data was split into various sets in order to remove any bias brought on by clustering and to prevent false high accuracy scores. Clustering was then carried out individually for each protein and each split. Then, depending on whether smooth translocations, interactions, or a combination of both should be the focus of the machine learning, signals from a certain cluster were chosen, and characteristics were then extracted from those signals. Following the under-sampling procedure, the data from several measurements were pooled. The test set was also treated in a similar fashion and the ML process was implemented as discussed earlier.

## AUTHOR CONTRIBUTIONS

**P.K.** initiated, conceptualized, and supervised the study. **S.D. and P.K.** developed the methodology. **H.Sh.** led the data analysis development with active participation from **S.D.** and **P.K. S.D.** and **H.Sh.** carried out the experimental measurements. Formal analysis was done by **H.Sh.** and **S.D.** with inputs from **P.K., B.K., and Y.M.N.D.Y.B. S.D.** wrote the original draft which was reviewed and edited by **P.K., H.Sh., B.K., Y.M.N.D.Y.B., E.D., and H.Su. H.Su.** and **E.D.** provided resources and strategies for the machine learning analysis. Funding was acquired by **P.K. and S.D.**

## COMPETING INTERESTS

The authors declare no competing interests.

## DATA AVAILABILITY

Data will be made available from the corresponding author upon request. The source code used for data extraction, clustering as well as machine learning analysis is hosted on GitHub: https://github.com/ssnl-anu/ProteinIdentifier

## ACKNOWLEDGEMENTS

This work used the ACT node of the NCRIS-enabled Australian National Fabrication Facility (ANFF-ACT). The authors acknowledge the NVIDIA Academic Hardware Grant Program for donation of two A6000 GPUs. S.D. was supported by an AINSE Ltd. Postgraduate Research Award (PGRA) and the Australian Government Research Training Program (RTP) Scholarship. P.K. acknowledges financial support from the Australian Research Council (ARC) under the ARC Discovery Project Scheme (DP180100068). This research was funded in part by and has been delivered in partnership with Our Health in Our Hands (OHIOH) – a strategic initiative of the Australian National University (ANU) – which aims to transform health care by developing new personalized health technologies and solutions in collaboration with patients, clinicians, and healthcare providers.

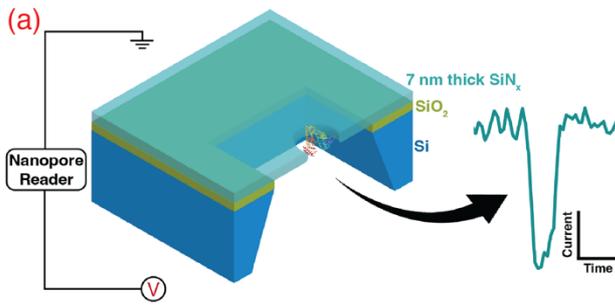
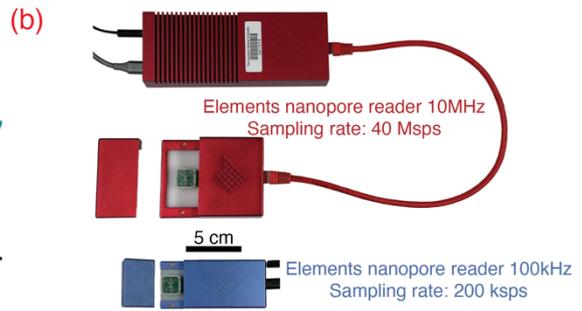
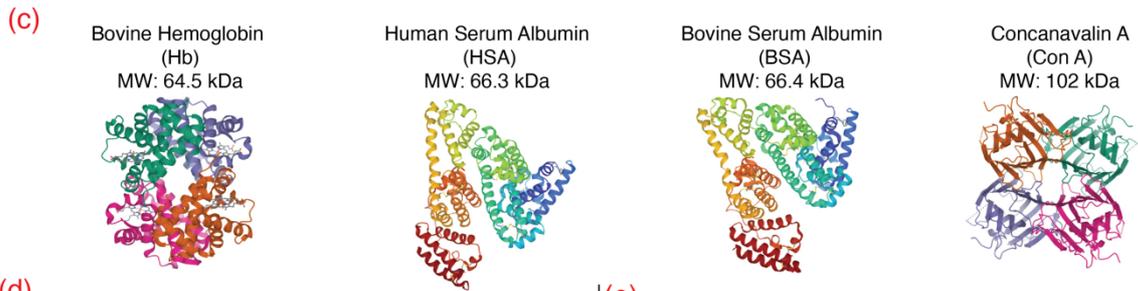
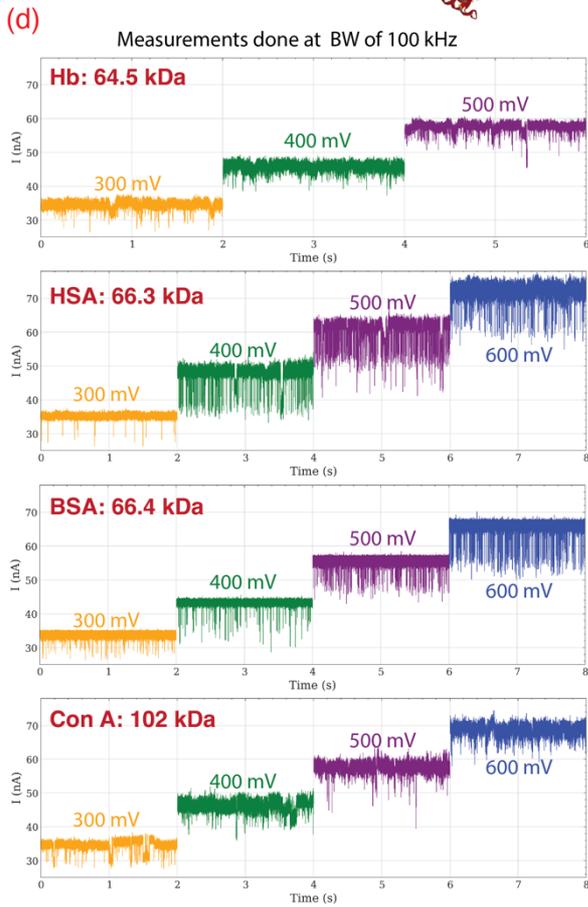
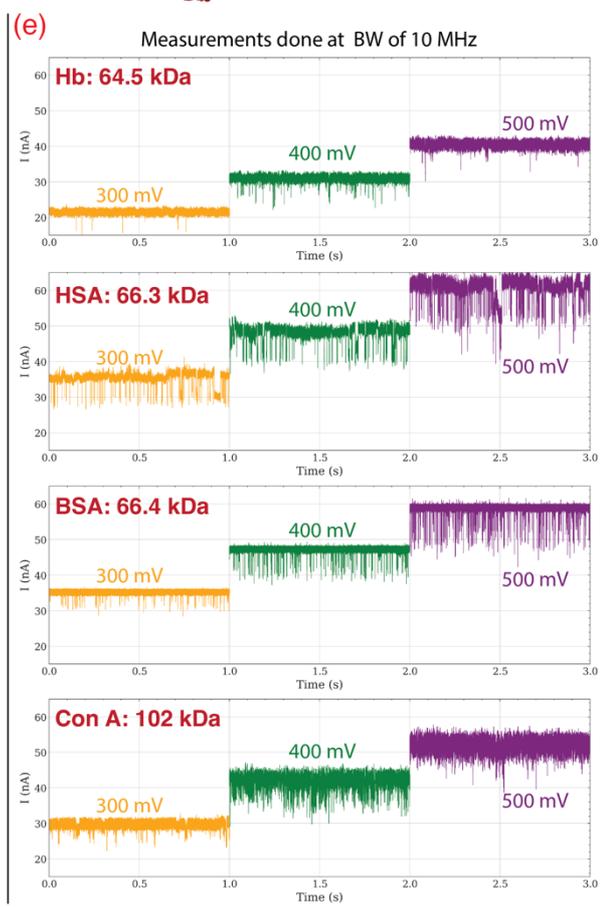
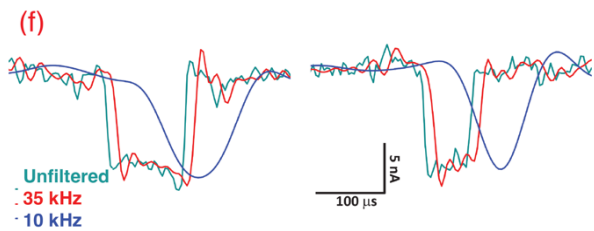
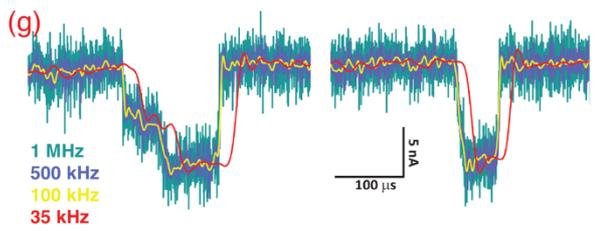

*Figure 1:* *(a) Sketch of protein translocation through a SiNx nanopore membrane. The proteins in a buffered electrolyte solution are placed in the cis and trans channels depending on the charge and polarity of the applied voltage. Silver-silver chloride electrodes are used to apply a bias across the membrane. (b) Photographs of the portable amplifiers used for data acquisition: Elements 10 MHz and 100 kHz nanopore readers providing a sampling rate of 40 Msps and 200 ksps, respectively. (c) Different proteins used in the present study. Protein structures were obtained from the Protein Data Bank website of the Research Collaboratory for Structural Bioinformatics. (d), (e) Representative ionic current-time traces of different proteins under bias dc voltage ranging from 300 mV to 500 mV recorded at sampling rates of 200 ksps and 40 Msps, respectively. (f), (g) Event shape transformation showing the loss of detail during low-pass filtering: comparison of the unfiltered signal and signals filtered at 35 kHz and 10 kHz frequencies (acquisition at 200 ksps) (f) and 1 MHz, 500 kHz, 100 kHz, and 35 kHz frequencies (acquisition at 40 Msps) (g). The measurements of Hb, HSA, BSA and Con A proteins were conducted in electrolyte solutions with pH of 7.94, 7.92, 8.02, and 8.02 respectively.*

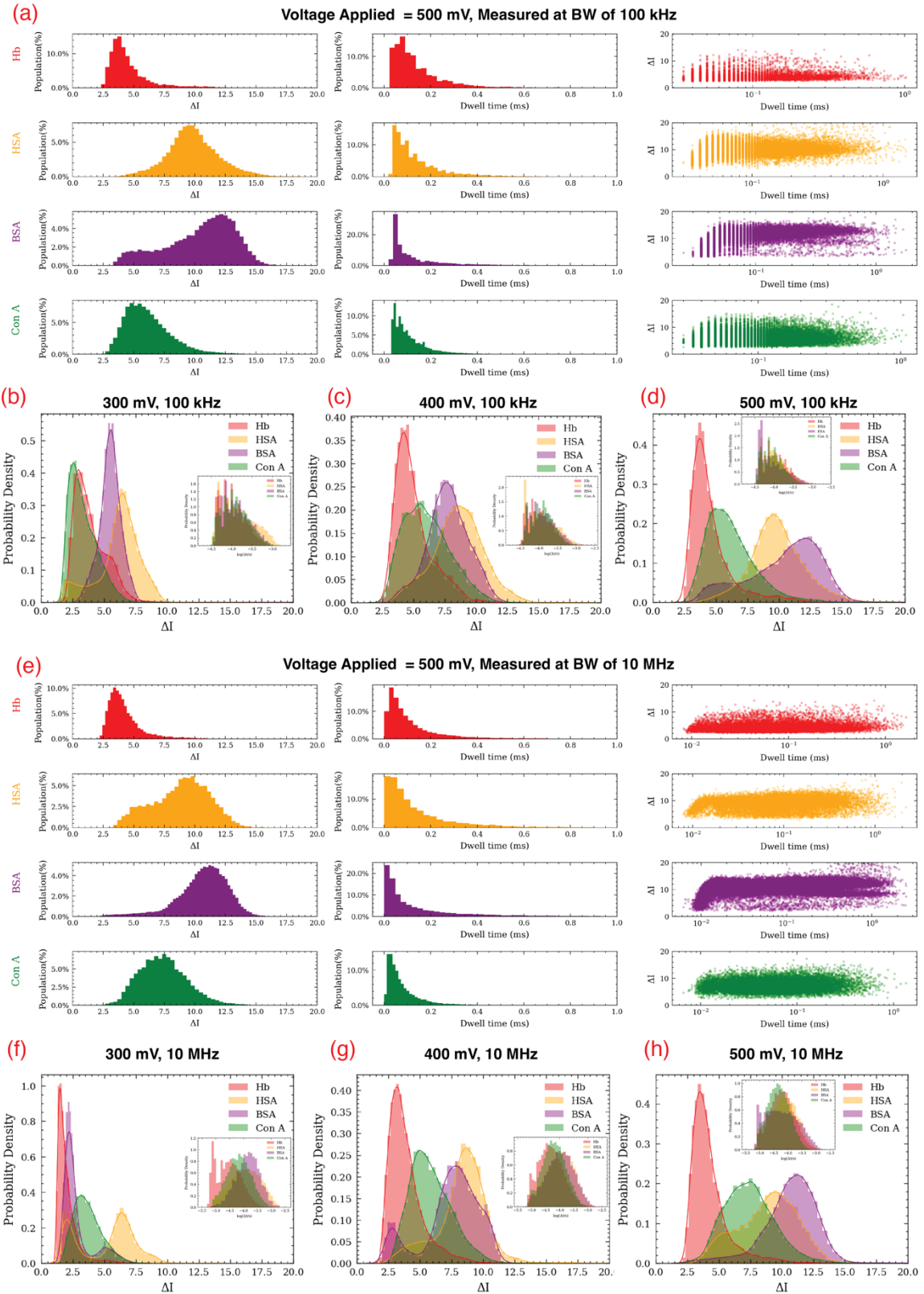

*Figure 2:* Histograms corresponding to the drop in ionic current (ΔI) and dwell time (Δt) as well as the scatter plots representing Δt as a function of ΔI for Hb, HSA, BSA, and Con A for measurements performed with nanopore readers of bandwidths 100 kHz (a) and 10 MHz (e) with a cross-membrane bias of 500mV. Overlap of the histograms corresponding to the change in relative conductance for measurements at 100 kHz BW when at a cross-membrane bias of 300 mV (b), 400 mV(c) and 500 mV(d) is applied. For measurements taken at 10 MHz BW, overlapping histograms are displayed in (f), (g), and (h). Overlapping histograms corresponding to the dwell time are shown in the insets of sub-figures (b-d) and (f-h).

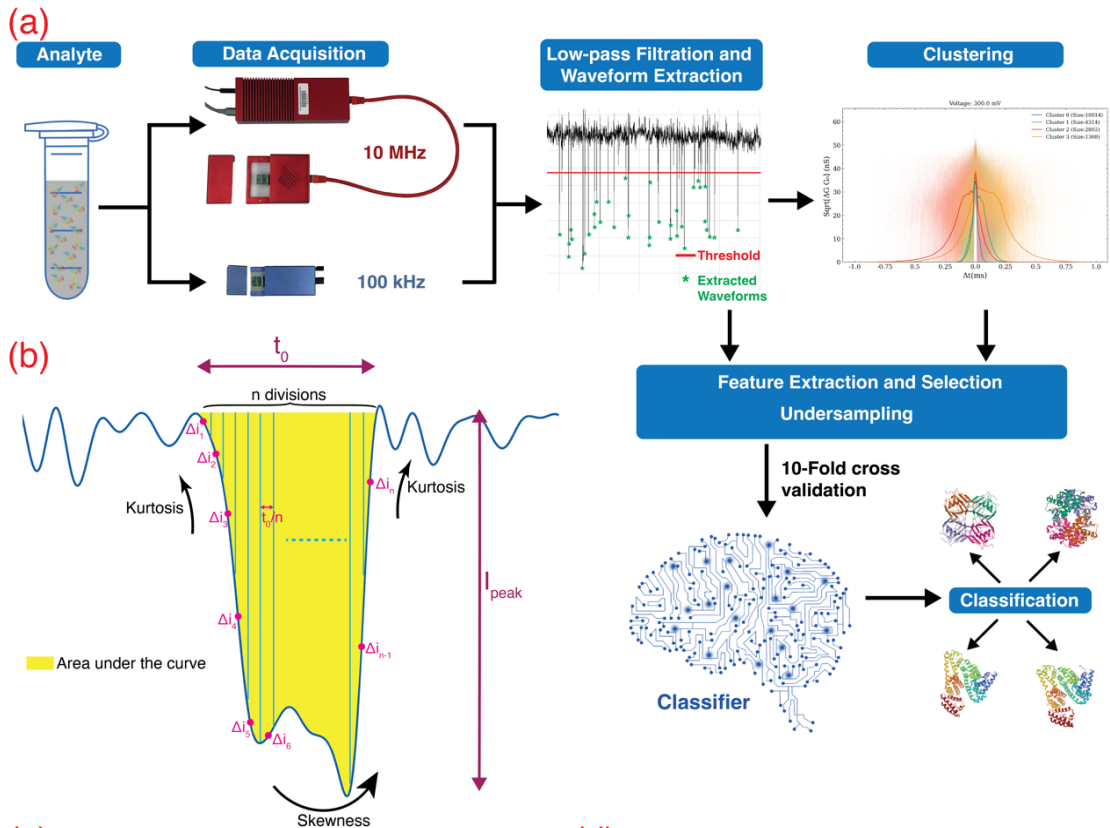

(a) Analyte → Data Acquisition (10 MHz, 100 kHz) → Low-pass Filtration and Waveform Extraction → Clustering → Feature Extraction and Selection Undersampling → 10-Fold cross validation → Classifier → Classification

(b) 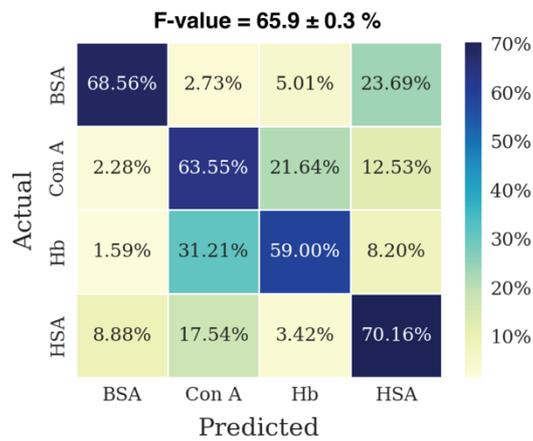

(c) 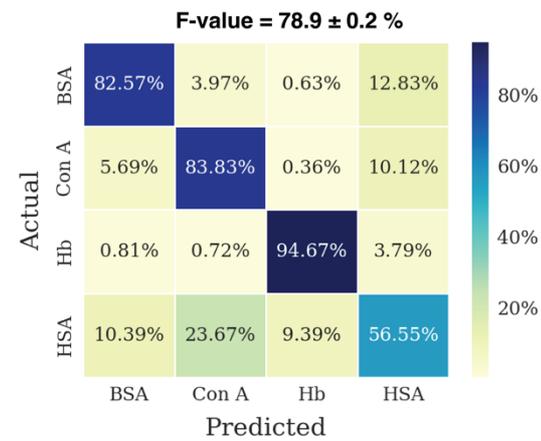

Voltage = 500 mV, BW = 100 kHz
F-value = 65.9 ± 0.3 %

(d) Voltage = 500 mV, BW = 10 MHz
F-value = 78.9 ± 0.2 %

(e) 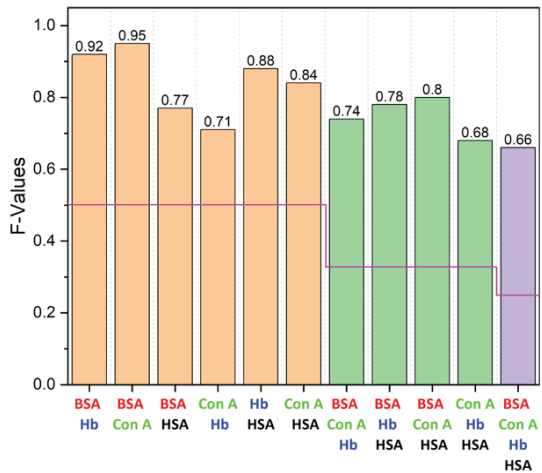

Voltage = 500 mV
Signals measured at BW of 100 kHz and filtered using 35 kHz LPF

(f) 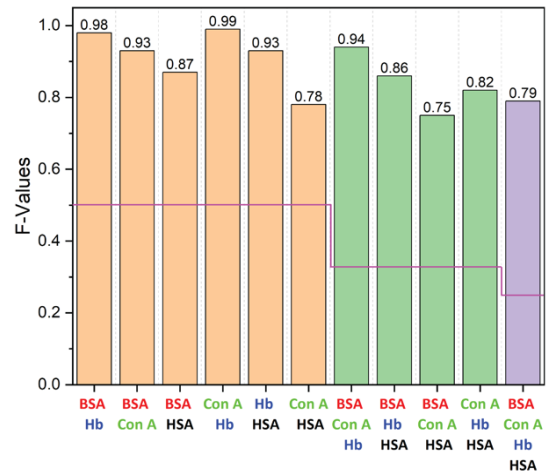

Voltage = 500 mV
Signals measured at BW of 10 MHz and filtered using 100 kHz LPF

*Figure 3: (a) Workflow for the label-free identification of proteins combining solid-state nanopores and machine learning. (b) Schematic displaying several resistive pulse feature parameters used for machine learning. Confusion matrices obtained for measurements taken at bandwidth of 100 kHz (c) and 10 MHz (d) using scheme 3 (cf. table 1). The applied bias was 500 mV. The darker the color in the matrix, the higher is the number of pulses corresponding to that combination. F-values for combinations of two, three and four different proteins employing measurements at bandwidths of 100 kHz (e) and 10 MHz (f). Orange represents combinations of two, green represents combinations of three, and purple shows combination of all four proteins. Under the same circumstances, measurements made at 40 Msps (BW = 10 MHz) result in a greater overall accuracy of identification of a single protein among four similarly sized proteins.*

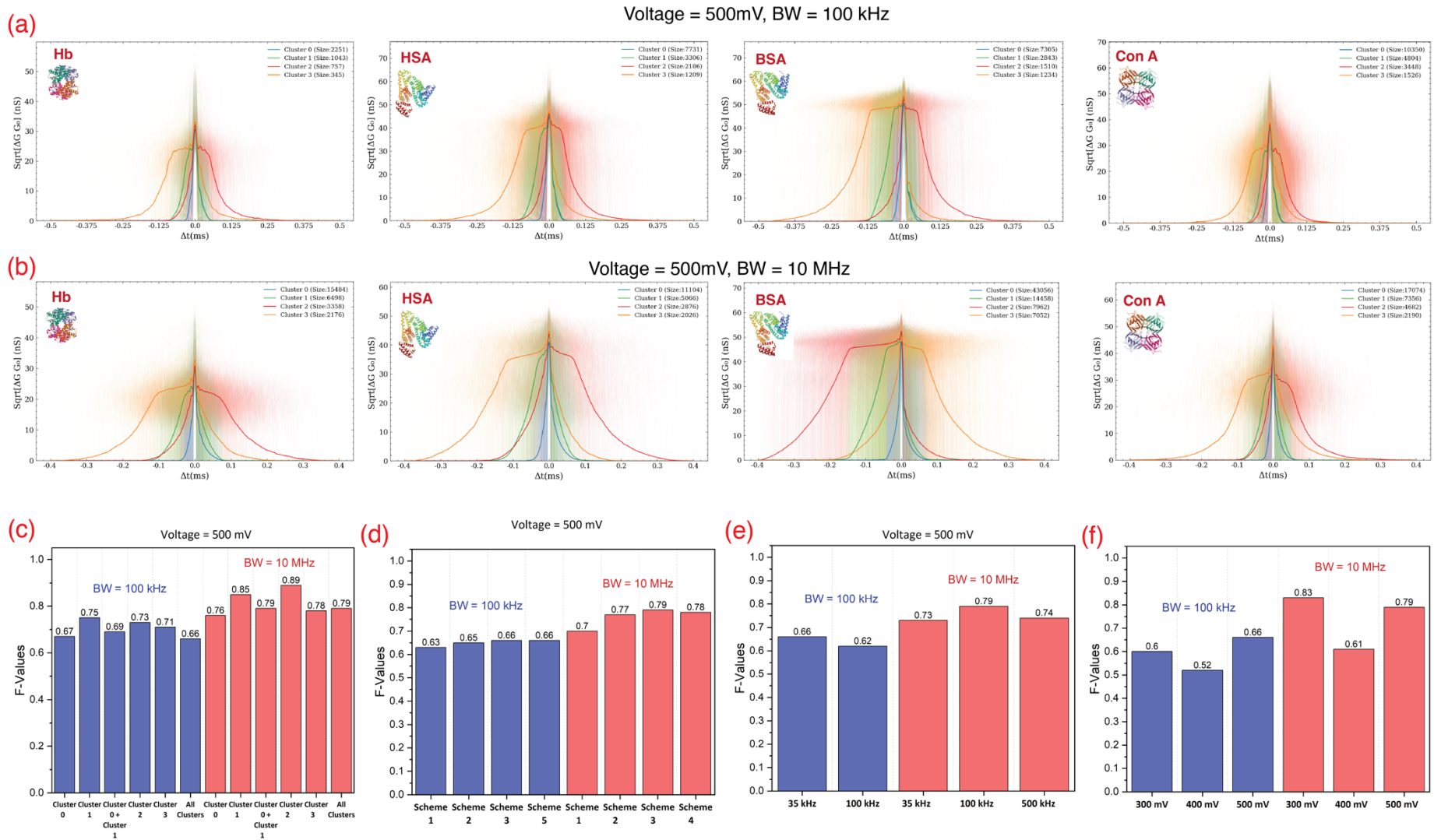

*Figure 4: Splitting the signals into different clusters to differentiate between possible events resulting from different conformations and interactions as a protein molecule translocates through a nanopore for measurements performed at bandwidth of (a) 100 kHz and (b) 10 MHz. The clusters shown are for*

*measurements with a trans-membrane bias of 500 mV. Due to surface charges on the nanopore wall and the proteins, specific and non-specific binding and electrostatic interactions can lead to such different signals (c) F-values (for the case of combination of all four proteins) as a function of events corresponding to different clusters used for training the classification model. 500 mV was applied, and feature extraction was performed using scheme 3. F-values (for the case of combination of all four proteins, applied voltage = 500 mV) as a function of extraction scheme (d) and the cutoff frequency of the low pass filter used to filter the raw data (e). (f) F-values as a function of applied voltage during the measurements. The F-values depicted in (c)-(f) are color coded, i.e., blue for the measurements done using nanopore reader with bandwidth of 100 kHz and red for measurements done using nanopore reader with bandwidth 10 MHz. The standard deviation of the calculated F-values is less than 0.009.*

*Table 1: Different feature extraction schemes and their corresponding features. The signal was divided into n parts of width $t_0/n$ and the average current drop corresponding to each part ($\Delta i_1$, $\Delta i_2$, ... $\Delta i_n$) was used as an input. Other features include current drop at full width at half maximum ($i_{fwhm}$), the peak drop in current ($i_{max}$), width at full width at half maximum ($t_{fwhm}$), area of the signal, skew and kurtosis. To account for the fluctuation in the nanopore size in different experiments, the values of the current characteristics were transformed into conductance and normalised by the open pore conductance.*

| Feature Extraction Scheme | Features/Details |
|---|---|
| Scheme 1 | $\Delta i_1, \Delta i_2, \ldots, \Delta i_{10} + t_0/10$ |
| Scheme 2 | $\Delta i_1, \Delta i_2, \ldots \Delta i_{10} + t_0/10 + t_{fwhm} + \text{area} + i_{fwhm} + i_{max}$ |
| Scheme 3 | $\Delta i_1, \Delta i_2, \ldots \Delta i_{10} + t_0/10 + t_{fwhm} + \text{area} + i_{fwhm} + i_{max} + \text{skew} + \text{kurtosis}$ |
| Scheme 4 | $\Delta i_1, \Delta i_2, \ldots \Delta i_{50} + t_0/50 + t_{fwhm} + \text{area} + i_{fwhm} + i_{max} + \text{skew} + \text{kurtosis}$ |
| Scheme 5 | Full signal + $t_{fwhm}$ + area + $i_{fwhm}$ + $i_{max}$ + skew + kurtosis |

# High Accuracy Protein Identification: Fusion of solid-state nanopore sensing and machine learning

Shankar Dutt[1, γ, *], Hancheng Shao[1, γ], Buddini Karawdeniya[2], Y. M. Nuwan D. Y. Bandara[3], Elena Daskalaki[4], Hanna Suominen[4,5], Patrick Kluth[1]

[1]*Department of Materials Physics, Research School of Physics, Australian National University, Canberra ACT 2601, Australia*

[2]*Department of Electronic Materials Engineering, Research School of Physics, Australian National University, Canberra ACT 2601, Australia*

[3]*Research School of Chemistry, Australian National University, Canberra ACT 2601, Australia*

[4]*School of Computing, College of Engineering, Computing and Cybernetics, Australian National University, Canberra ACT 2601, Australia*

[5]*Eccles Institute of Neuroscience, College of Health and Medicine, Australian National University, Canberra ACT 2601, Australia*

γ: These authors contributed equally.

*Email: shankar.dutt@anu.edu.au

# TABLE OF CONTENTS

| Figure S6 | Overlap of histograms corresponding to the dwell time for measurements done at 40 Msps (BW = 10 MHz) when the applied bias across the membrane was 300 mV (a), 400 mV (b) and 500 mV (c). |
|---|---|
| Figure S7 | Clustering of the signals to differentiate possible translocation events from one another as Hb translocates through a nanopore. The clusters are shown for measurements done at 200 ksps (BW = 100 kHz) under an applied bias of 300 mV (a) and 400 mV (b). Also shown are clusters resulting from measurements done at 40 Msps (BW = 10 MHz) under an applied bias of 300 mV (c) and 400 mV (d). |
| Figure S8 | Clustering of the signals to differentiate possible translocation events from one another as HSA translocates through a nanopore. The clusters are shown for measurements done at 200 ksps (BW = 100 kHz) under an applied bias of 300 mV (a) and 400 mV (b). Also shown are clusters resulting from measurements done at 40 Msps (BW = 10 MHz) under an applied bias of 300 mV (c) and 400 mV (d). |
| Figure S9 | Clustering of the signals to differentiate possible translocation events from one another as BSA translocates through a nanopore. The clusters are shown for measurements done at 200 ksps (BW = 100 kHz) under an applied bias of 300 mV (a) and 400 mV (b). Also shown are clusters resulting from measurements done at 40 Msps (BW = 10 MHz) under an applied bias of 300 mV (c) and 400 mV (d). |
| Figure S10 | Clustering of the signals to differentiate possible translocation events from one another as con A translocates through a nanopore. The clusters are shown for measurements done at 200 ksps (BW = 100 kHz) under an applied bias of 300 mV (a) and 400 mV (b). Also shown are clusters resulting from measurements done at 40 Msps (BW = 10 MHz) under an applied bias of 300 mV (c) and 400 mV (d). |
| Figure S11 | Confusion matrix from machine learning analysis. |
| Figure S12 | Confusion matrices of 2 types of proteins at 300mV using 200 ksps data (BW = 100 kHz) and 40 Msps data (BW = 10 MHz) employing Scheme 3. The numbers in the matrix element indicate the number of waveforms corresponding to that combination while the color of the element indicates the accuracy. The darker the color the higher the accuracy. |

| Figure S13 | Confusion matrices of 2 types of proteins at 400mV using 200 ksps data (BW = 100 kHz) and 40 Msps data (BW = 10 MHz) employing Scheme 3. The numbers in the matrix element indicate the number of waveforms corresponding to that combination while the color of the element indicates the accuracy. The darker the color the higher the accuracy. |
|---|---|
| Figure S14 | Confusion matrices of 2 types of proteins at 500mV using 200 ksps data (BW = 100 kHz) and 40 Msps data (BW = 10 MHz) employing Scheme 3. The numbers in the matrix element indicate the number of waveforms corresponding to that combination while the color of the element indicates the accuracy. The darker the color the higher the accuracy. |
| Figure S15 | Confusion matrices of 2 types of proteins at 600 mV using 200 ksps data (BW = 100 kHz) employing Scheme 3. The numbers in the matrix element indicate the number of waveforms corresponding to that combination while the color of the element indicates the accuracy. The darker the color the higher the accuracy. |
| Figure S16 | Confusion matrices of 3 types of proteins at 300mV using 200 ksps data (BW = 100 kHz) and 40 Msps data (BW = 10 MHz) employing Scheme 3. The numbers in the matrix element indicate the number of waveforms corresponding to that combination while the color of the element indicates the accuracy. The darker the color the higher the accuracy. |
| Figure S17 | Confusion matrices of 3 types of proteins at 400mV using 200 ksps data (BW = 100 kHz) and 40 Msps data (BW = 10 MHz) employing Scheme 3. The numbers in the matrix element indicate the number of waveforms corresponding to that combination while the color of the element indicates the accuracy. The darker the color the higher the accuracy. |
| Figure S18 | Confusion matrices of 3 types of proteins at 500mV using 200 ksps data (BW = 100 kHz) and 40 Msps data (BW = 10 MHz) employing Scheme 3. The numbers in the matrix element indicate the number of waveforms corresponding to that combination while the color of the element indicates the accuracy. The darker the color the higher the accuracy. |
| Figure S19 | Confusion matrices of 3 types of proteins at 600mV using 200 ksps data (BW = 100 kHz) employing Scheme 3. The numbers in the matrix element indicate the number of waveforms corresponding to that combination while |
| | the color of the element indicates the accuracy. The darker the color the higher the accuracy. |
| **Figure S20** | Confusion matrices of 4 types of proteins at 300mV, 400mV and 500mV using 200 ksps data (BW = 100 kHz) and 40 Msps data (BW = 10 MHz) employing Scheme 3. The numbers in the matrix element indicate the number of waveforms corresponding to that combination while the color of the element indicates the accuracy. The darker the color the higher the accuracy. |
| **Figure S21** | F-values in combinations of two, three and four types of proteins when data extraction was done using a lowpass filter of (a) 35 kHz and (b) 100 kHz for data acquired at 200 ksps (BW = 100 kHz). The feature extraction was performed employing scheme 3. |
| **Figure S22** | F- values in combinations of two, three and four types of proteins when data extraction was done using a lowpass filter of (a) 35 kHz, (b) 100 kHz and (c) 500 kHz for data acquired at 40 Msps. The feature extraction was performed employing scheme 3. |

*Table S1:* *Details of proteins used in this study. The approximate values for the isoelectric points (pI) were obtained from the manufacturer or from other easily accessible data sources.*

| Protein | Product Details | MW | Isoelectric point |
|---|---|---|---|
| Bovine Hemoglobin (Hb) | Sigma Aldrich, H2500 | 64.5 kDa | 7.1 |
| Human Serum Albumin (HAS) | Sigma Aldrich, A9511 | 66.3 kDa | 4.7 |
| Bovine Serum Albumin (BSA) | Sigma Aldrich, A2153 | 66.4 kDa | 4.6 |
| Concanavalin A (Con A) | Sigma Aldrich, C2010 | 102 kDa | 4.9 |

**Table S22:** *Characteristics (open pore conductance ($G_0$) using 1 M KCl, pH ~7, calculated pore diameter and protein measurement done) of the nanopores used to measure the proteins. Results in the main text are shown corresponding to Chip ID of CHN139, CHN122, CHN121 and CHN118.*

| Chip ID | Protein Measured | $G_0$ (nS) | Calculated Pore Diameter |
|---|---|---|---|
| CHN139 | Hb | 111 ± 2 | 16.1 ± 0.4 |
| CHN119 | Hb | 105 ± 1 | 15.5 ± 0.3 |
| CHN122 | HSA | 103 ± 3 | 15.3 ± 0.4 |
| CHN121 | Con A | 116 ± 1 | 16.6 ± 0.3 |
| CHN72 | Con A | 102 ± 1 | 15.1 ± 0.3 |
| CHN118 | BSA | 117 ± 2 | 16.8 ± 0.4 |
| CHN64 | BSA | 105 ± 1 | 15.5 ± 0.3 |
| CHN116 | BSA | 104 ± 2 | 15.4 ± 0.4 |

*Table S3*: *Number of extracted signals for different measurements at different voltages for different proteins corresponding to Chip ID of CHN139, CHN122, CHN121 and CHN118.*

| Protein | Voltage Applied (mV) | Number of waveforms extracted (measured at 200 ksps, BW = 100 kHz) | Number of waveforms extracted (measured at 40 Msps, BW = 10 MHz) |
|---|---|---|---|
| Hb | 300 | 18,003 | 32,149 |
| Hb | 400 | 4,330 | 57,097 |
| Hb | 500 | 8,320 | 17,504 |
| HSA | 300 | 39,552 | 64,310 |
| HSA | 400 | 27,325 | 32,644 |
| HSA | 500 | 39,345 | 24,810 |
| HSA | 600 | 40,580 | - |
| BSA | 300 | 7,605 | 8,034 |
| BSA | 400 | 19,563 | 22,653 |
| BSA | 500 | 29,509 | 62,066 |
| BSA | 600 | 24,186 | - |
| Con A | 300 | 35,908 | 16,778 |
| Con A | 400 | 18,992 | 21,095 |
| Con A | 500 | 58,312 | 21,925 |
| Con A | 600 | 40,004 | - |

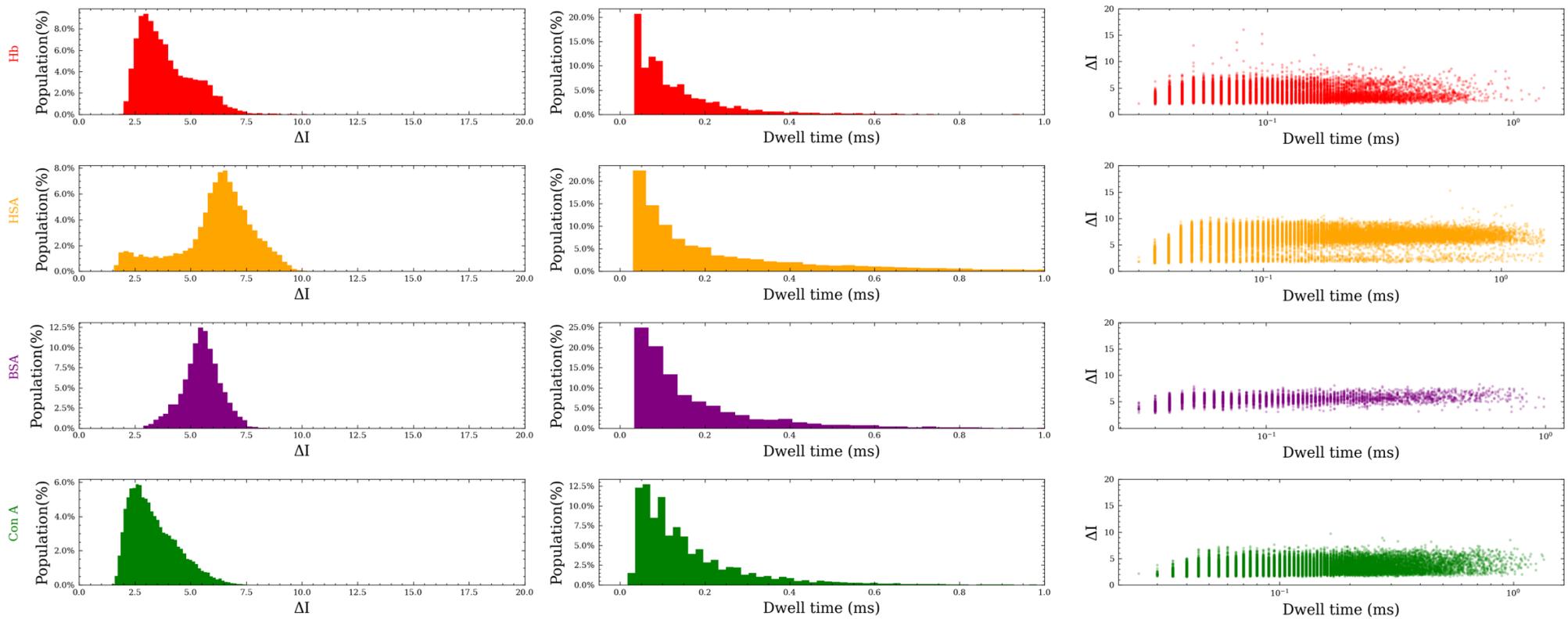

*Figure S1*: *Histograms corresponding to the drop in ionic current (ΔI), dwell time (Δt) as well as the scatter plots representing Δt as a function of ΔI for Hb, HSA, BSA, and Con A for measurements done at 200 ksps (BW = 100 kHz) for an applied voltage of 300 mV across the membrane.*

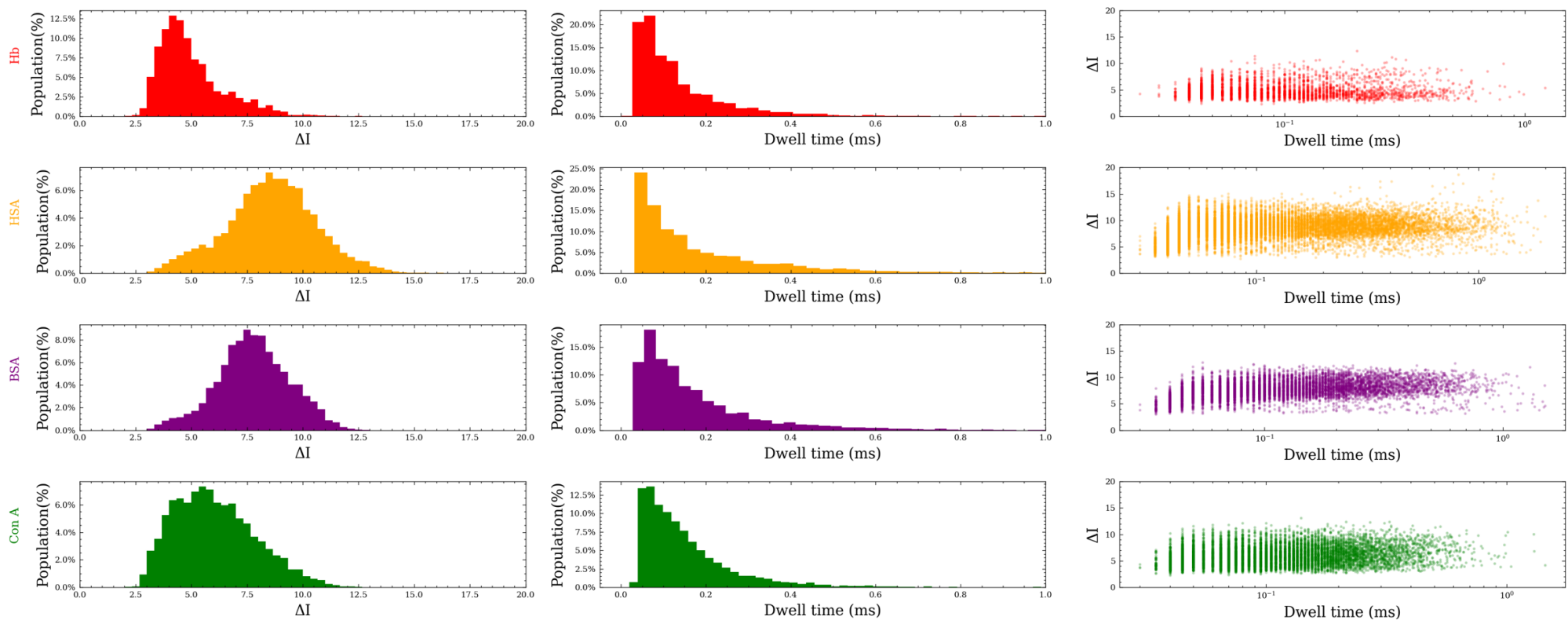

***Figure S2:*** *Histograms corresponding to the drop in ionic current (ΔI), dwell time (Δt) as well as the scatter plots representing Δt as a function of ΔI for Hb, HSA, BSA, and Con A for measurements done at 200 ksps (BW = 100 kHz) for an applied voltage of 400 mV across the membrane.*

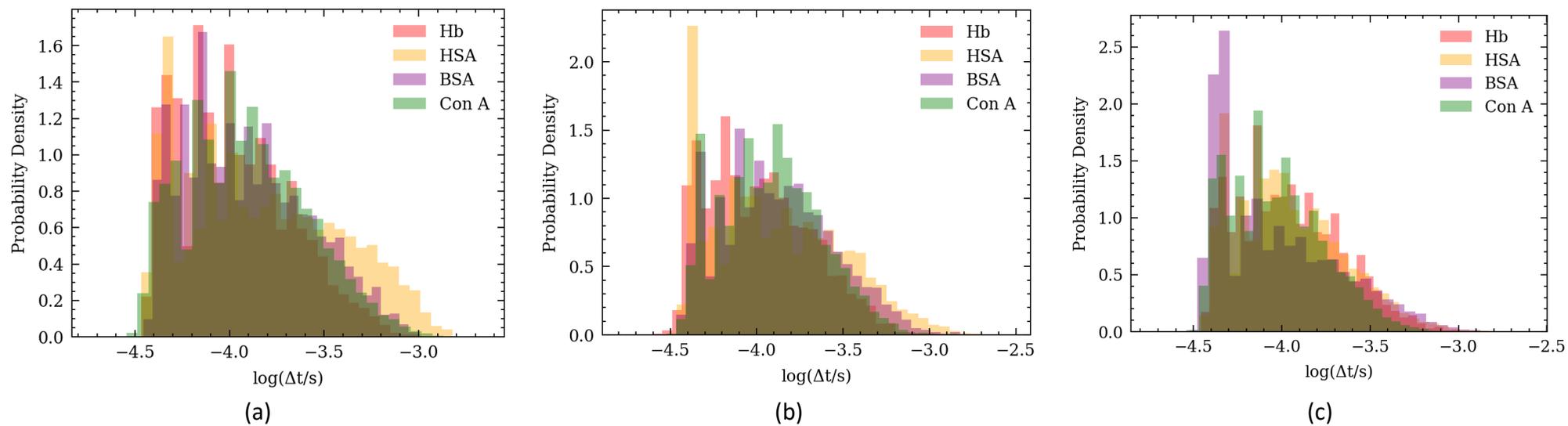

*Figure S3:* *Overlap of the histograms corresponding to the dwell time for measurements done at 200 ksps (BW = 100 kHz) when the applied bias across the membrane was 300 mV (a), 400 mV(b) and 500 mV(c).*

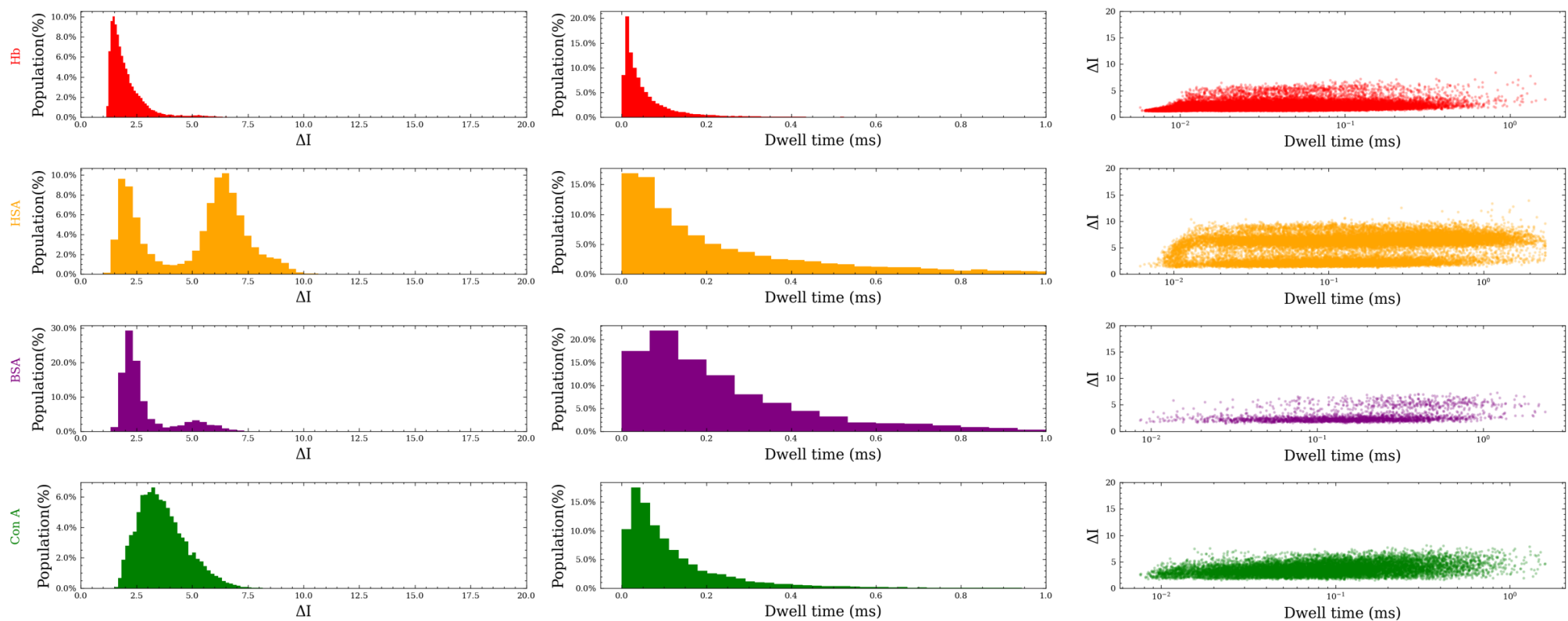

*Figure S4*: Histograms corresponding to the drop in ionic current (ΔI), dwell time (Δt) as well as the scatter plots representing Δt as a function of ΔI for Hb, HSA, BSA, and Con A for measurements done at 40 Msps (BW = 10 MHz) for an applied voltage of 300 mV across the membrane.

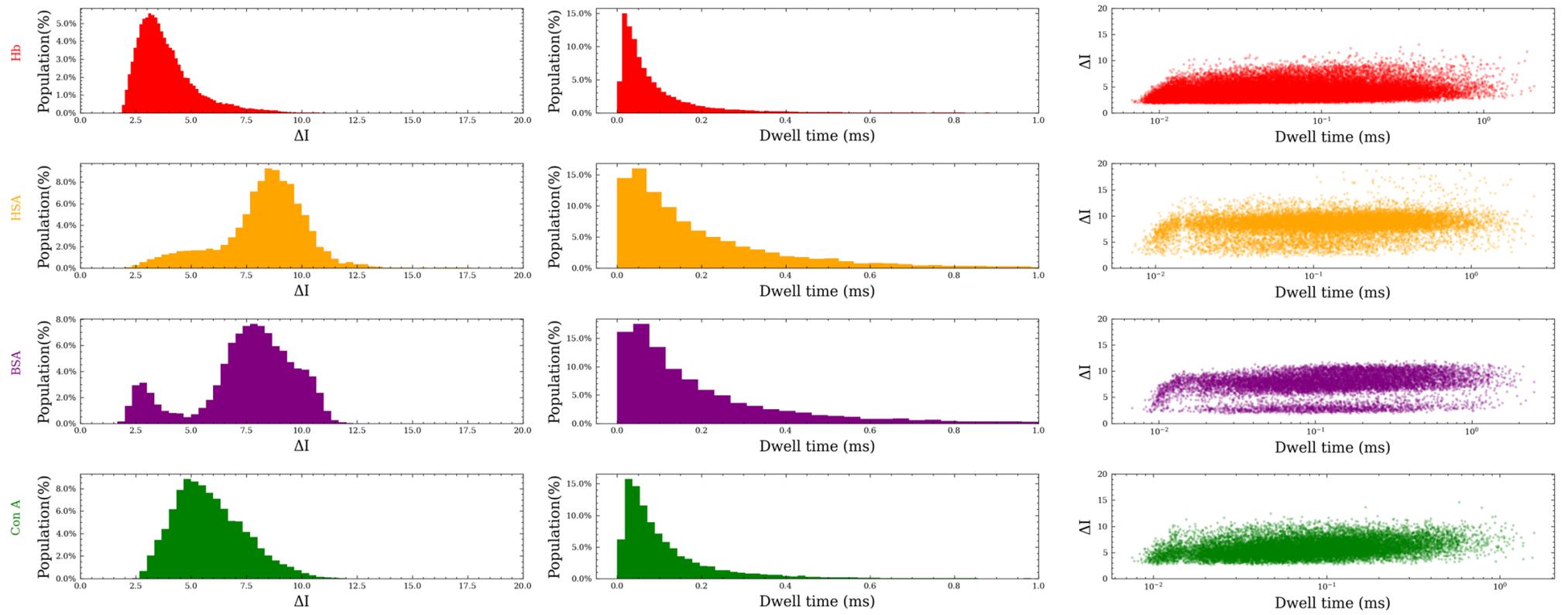

***Figure S5:*** *Histograms corresponding to the drop in ionic current (ΔI), dwell time (Δt) as well as the scatter plots representing Δt as a function of ΔI for Hb, HSA, BSA, and Con A for measurements done at 40 Msps (BW = 10 MHz) for an applied voltage of 400 mV across the membrane.*

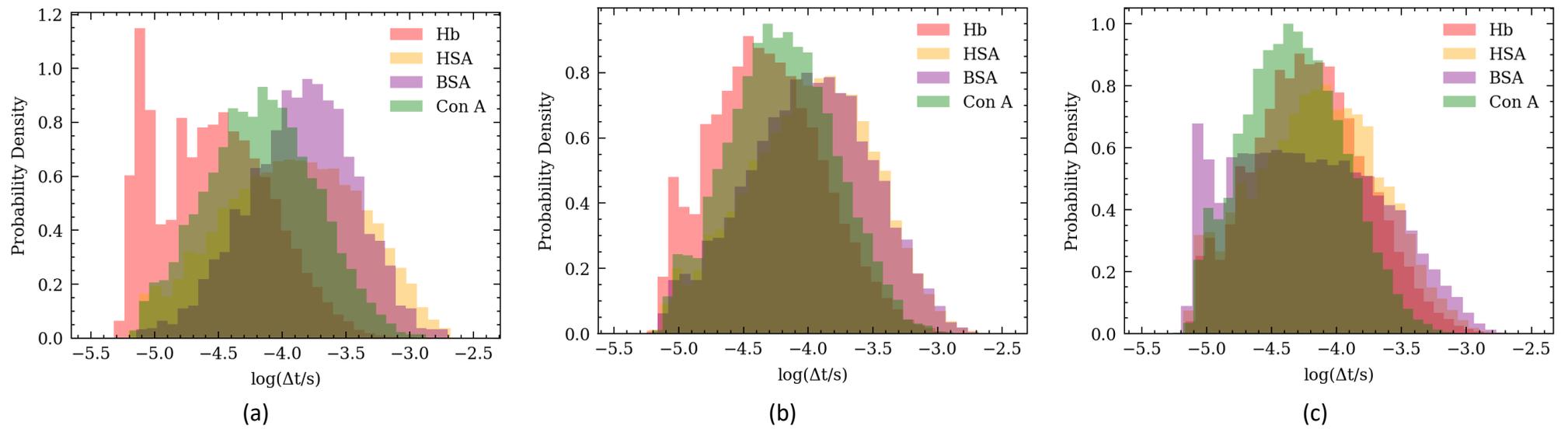

***Figure S6:*** *Overlap of histograms corresponding to the dwell time for measurements done at 40 Msps (BW = 10 MHz) when the applied bias across the membrane was 300 mV (a), 400 mV (b) and 500 mV (c).*

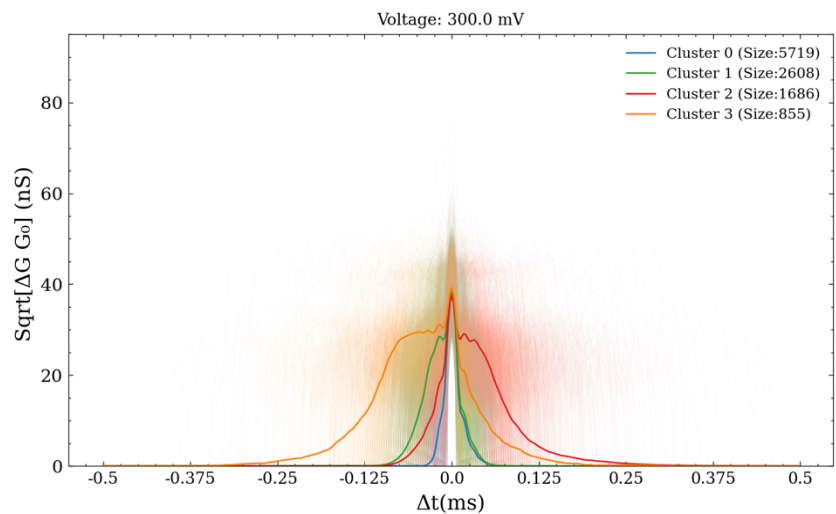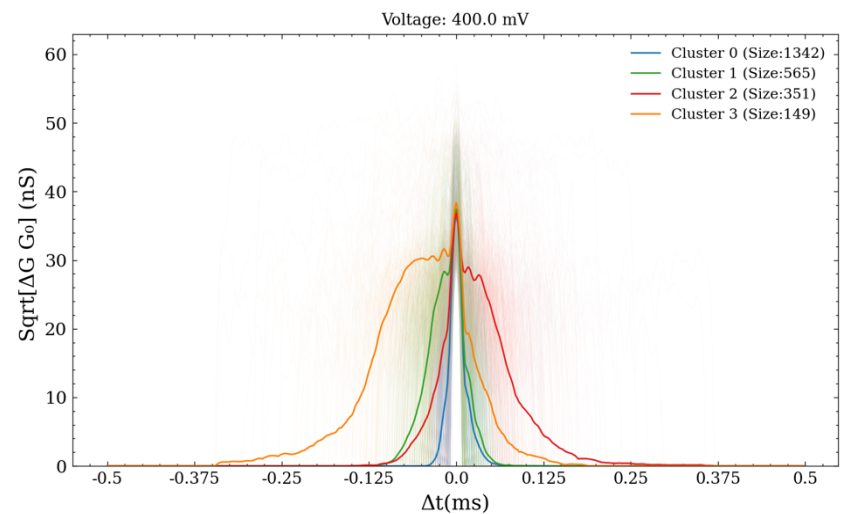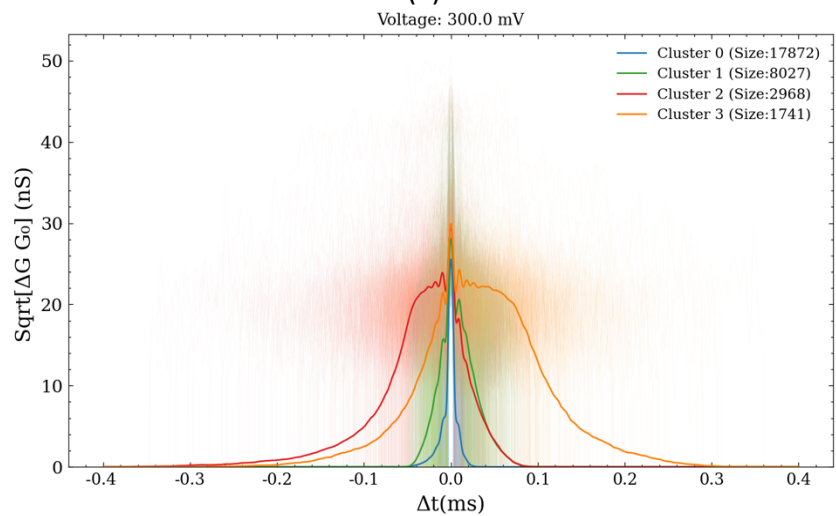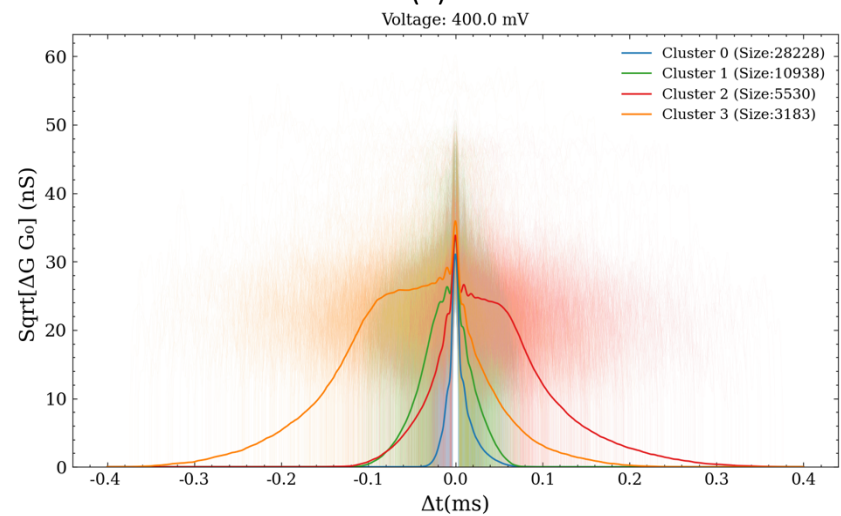

*Figure S7:* Clustering of the signals to differentiate possible translocation events from one another as Hb translocates through a nanopore. The clusters are shown for measurements done at 200 ksps (BW = 100 kHz) under an applied bias of 300 mV (a) and 400 mV (b). Also shown are clusters resulting from measurements done at 40 Msps (BW = 10 MHz) under an applied bias of 300 mV (c) and 400 mV (d).

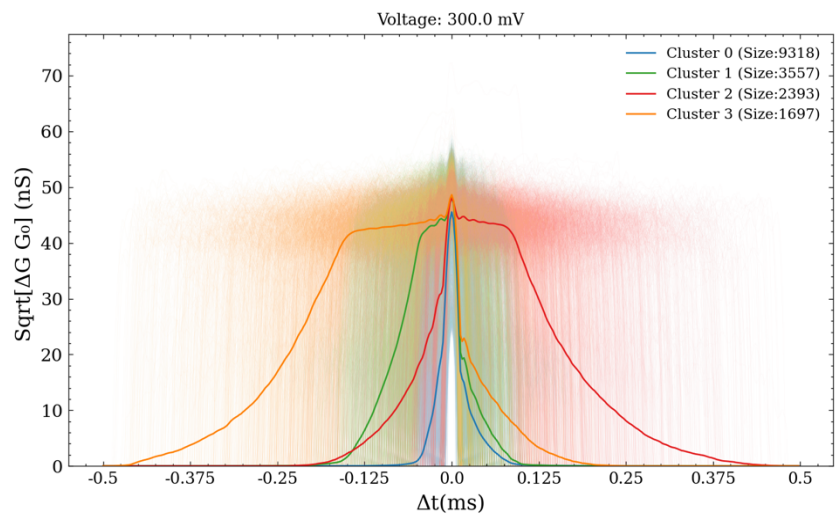
(a)

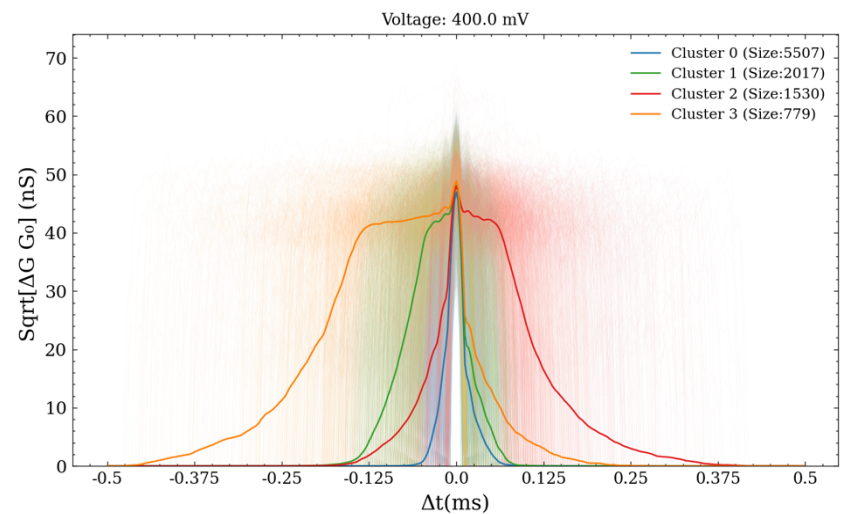
(b)

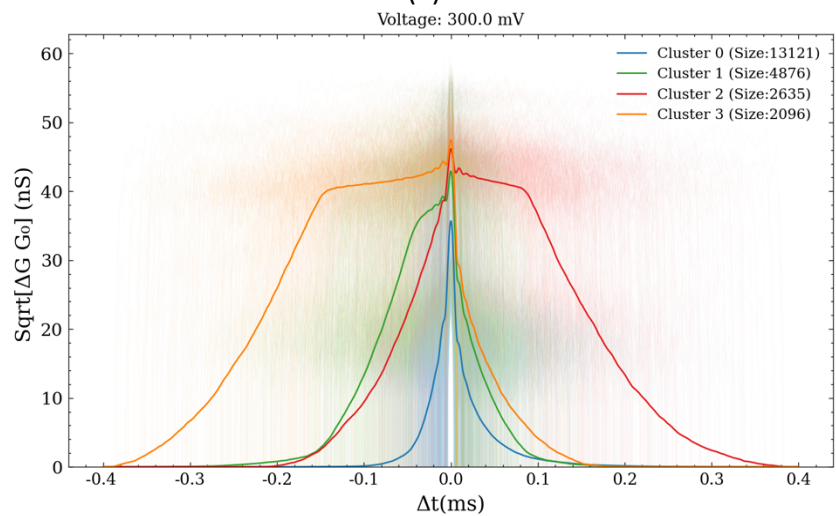
(c)

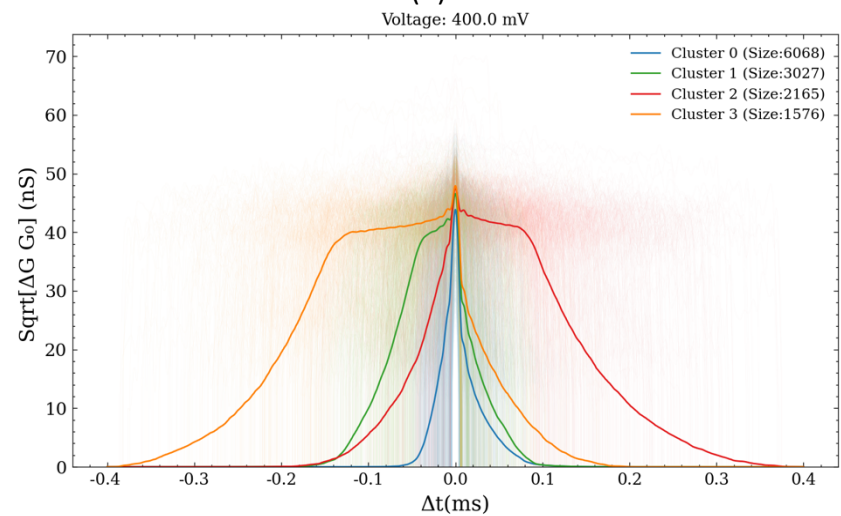
(d)

***Figure S8:*** *Clustering of the signals to differentiate possible translocation events from one another as HSA translocates through a nanopore. The clusters are shown for measurements done at 200 ksps (BW = 100 kHz) under an applied bias of 300 mV (a) and 400 mV (b). Also shown are clusters resulting from measurements done at 40 Msps (BW = 10 MHz) under an applied bias of 300 mV (c) and 400 mV (d).*

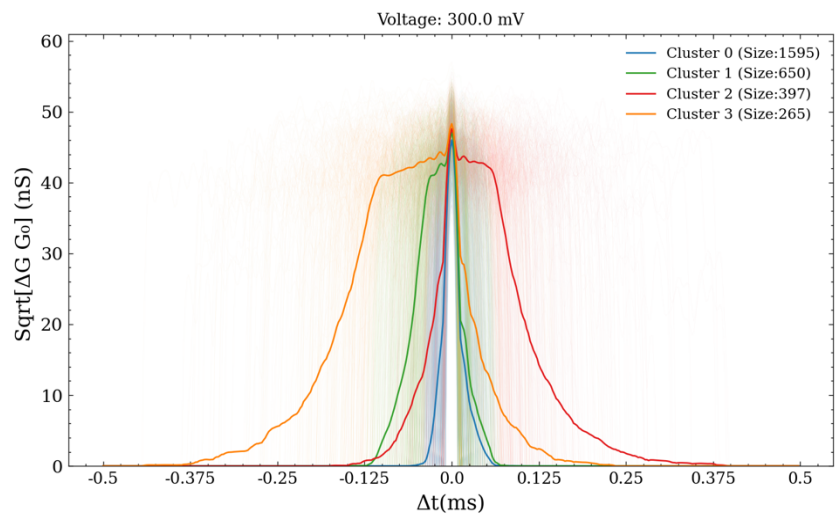
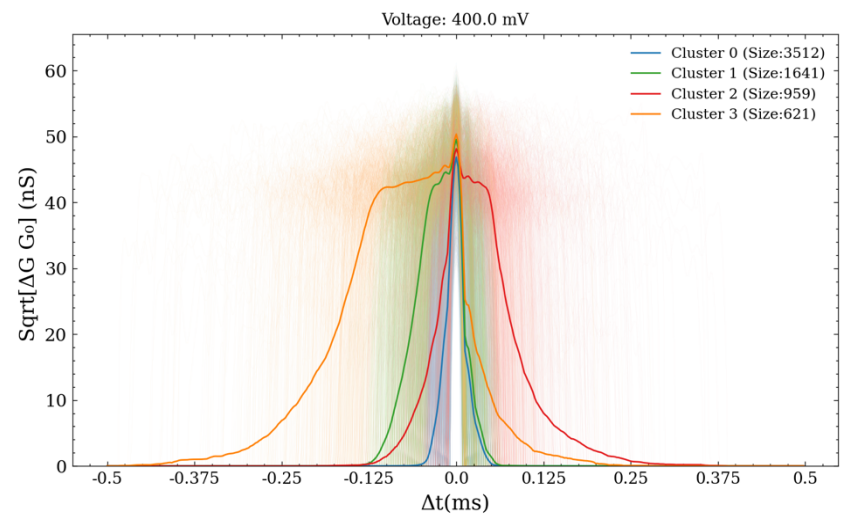
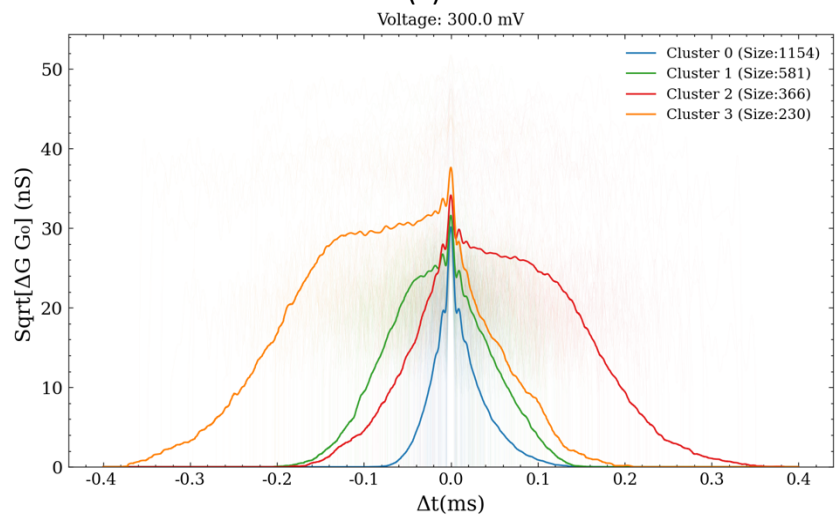
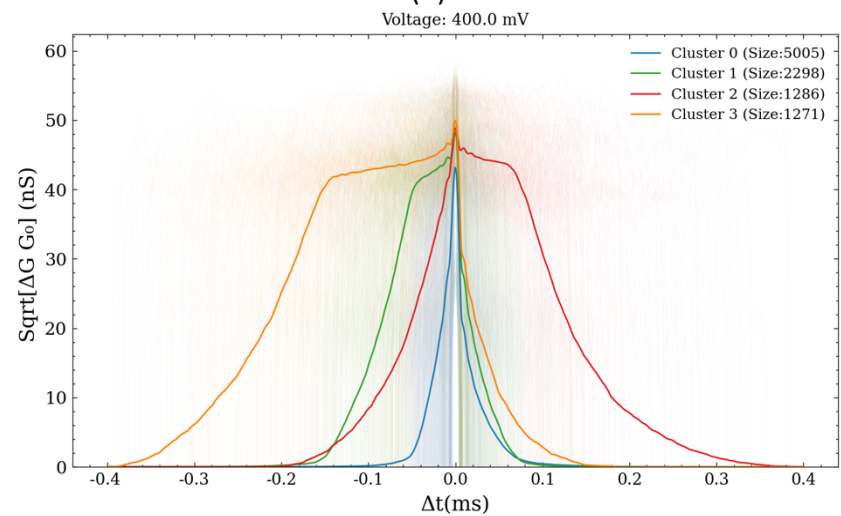

*Figure S9:* Clustering of the signals into to differentiate possible translocation events from one another as BSA translocates through a nanopore. The clusters are shown for measurements done at 200 ksps (BW = 100 kHz) under an applied bias of 300 mV (a) and 400 mV (b). Also shown are clusters resulting from measurements done at 40 Msps (BW = 10 MHz) under an applied bias of 300 mV (c) and 400 mV (d).

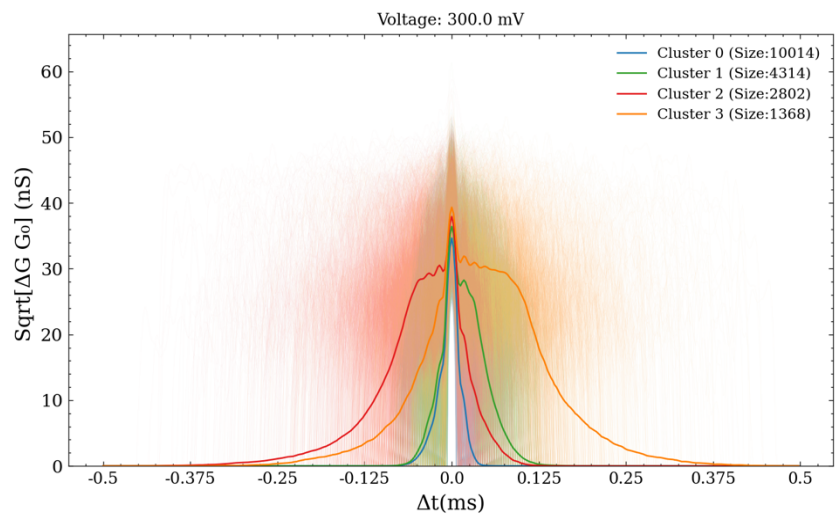
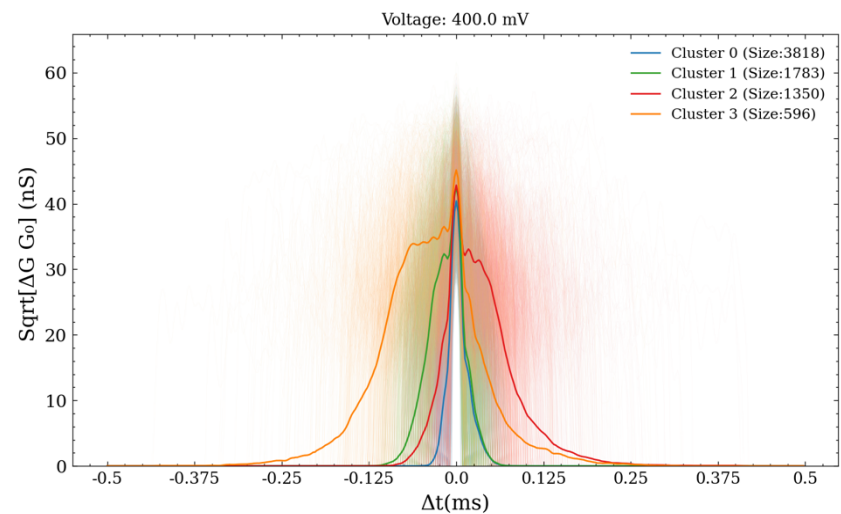
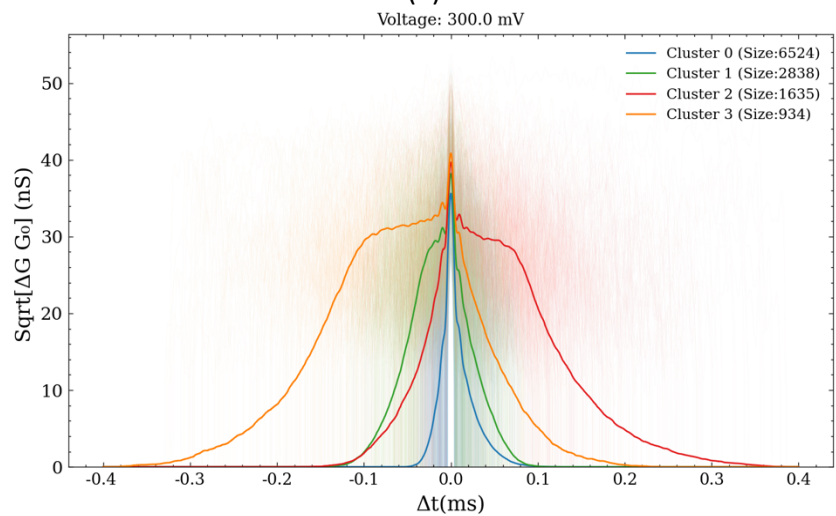
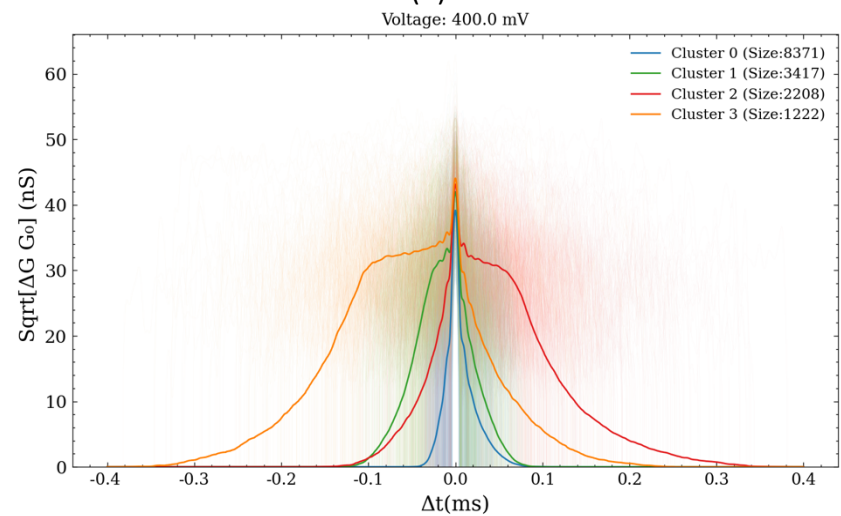

***Figure S10:*** *Clustering of the signals to differentiate possible translocation events from one another as con A translocates through a nanopore. The clusters are shown for measurements done at 200 ksps (BW = 100 kHz) under an applied bias of 300 mV (a) and 400 mV (b). Also shown are clusters resulting from measurements done at 40 Msps (BW = 10 MHz) under an applied bias of 300 mV (c) and 400 mV (d).*

|              | TP | FP |
|--------------|----|----|
| **Actual**   | FP | TN |
|              | **Predicted** | |

*Figure S111:* Confusion matrix from machine learning analysis. Ratio of True Positive (TP), True Negative (TN), False Positive (FP) and False Negative (FN) defines Precision, Recall/Sensitivity, Specificity and F-value as follows:

$$Precision = \frac{TP}{TP + FP}$$

$$\text{Recall/Sensitivity} = \frac{TP}{TP + FN}$$

$$Specificity = \frac{TN}{TN + FP}$$

$$F-value = \frac{2 \times Recall \times Precision}{Recall + Precision}$$

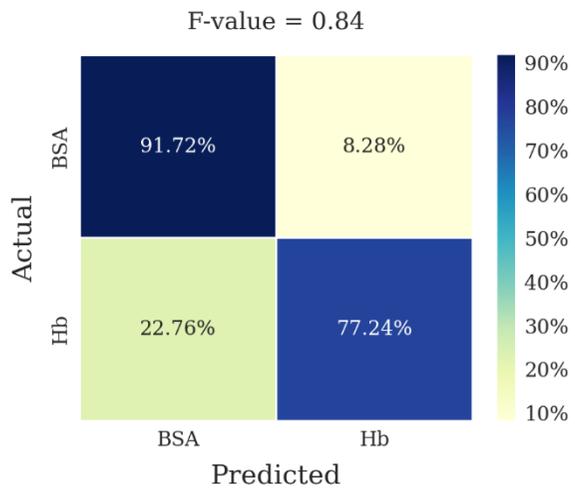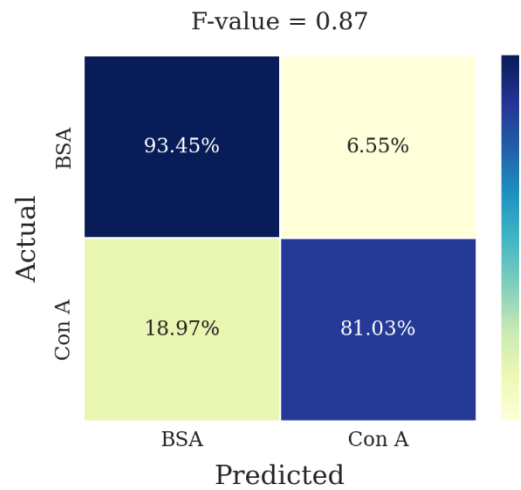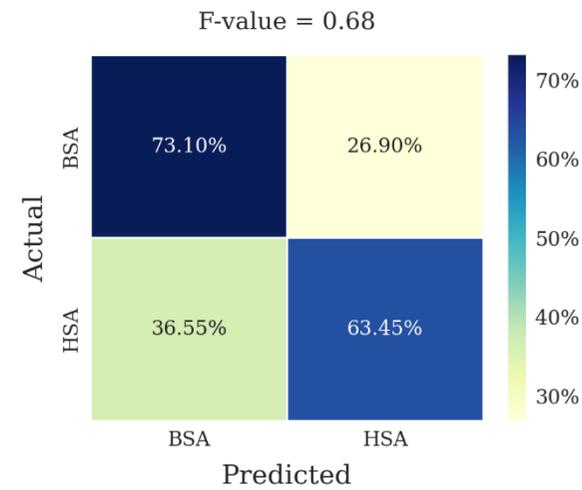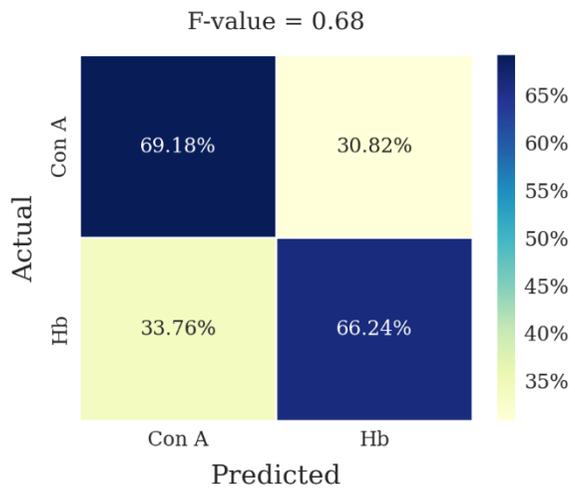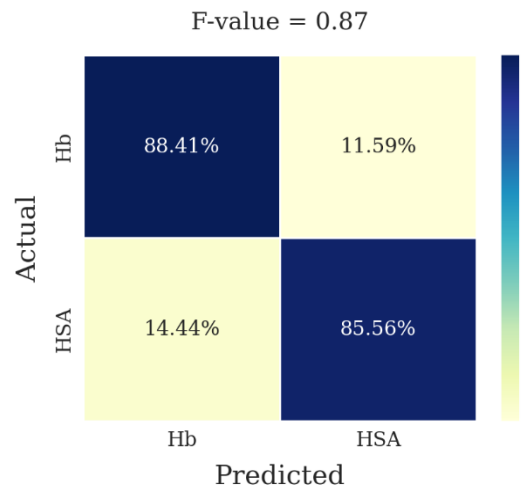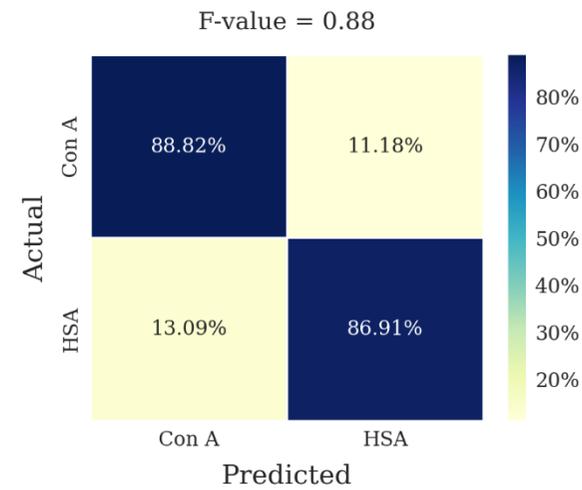

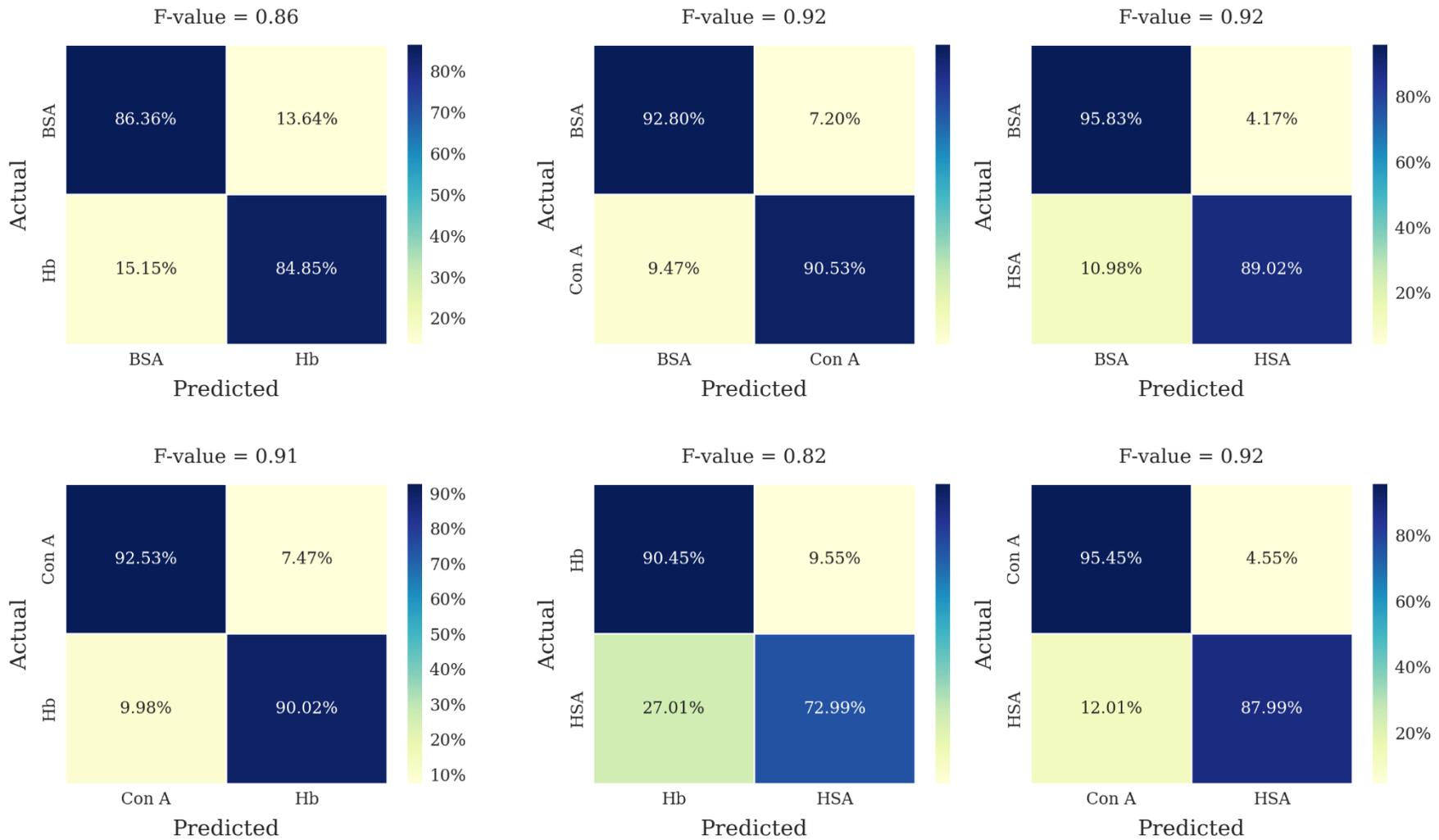

*Figure S12:* Confusion matrices of 2 types of proteins at 300mV using 200 ksps data (BW = 100 kHz) and 40 Msps data (BW = 10 MHz) employing Scheme 3. The numbers in the matrix element indicate the number of waveforms corresponding to that combination while the color of the element indicates the accuracy. The darker the color the higher the accuracy.

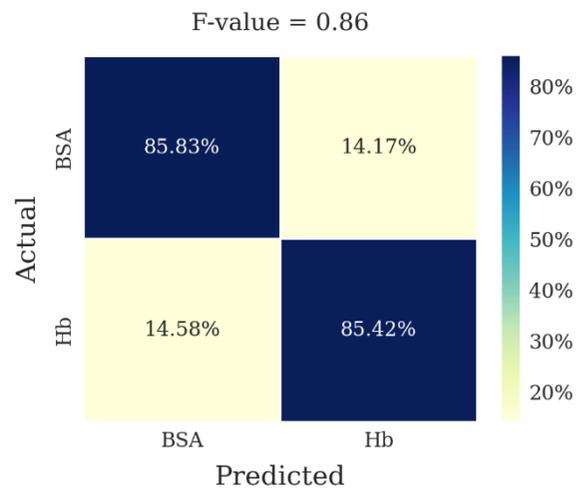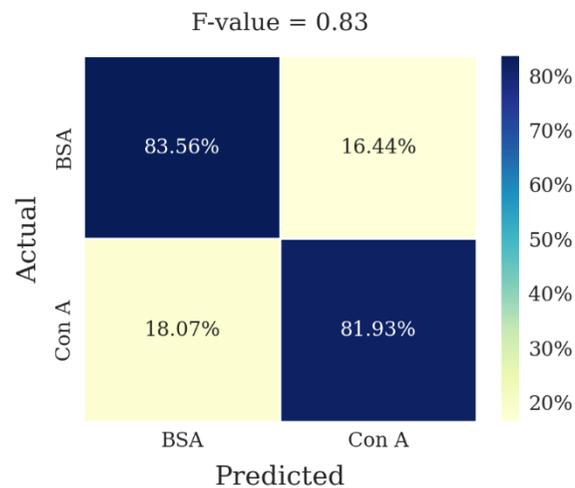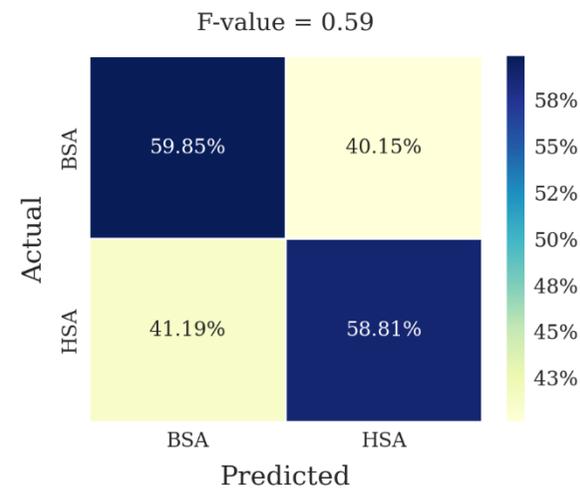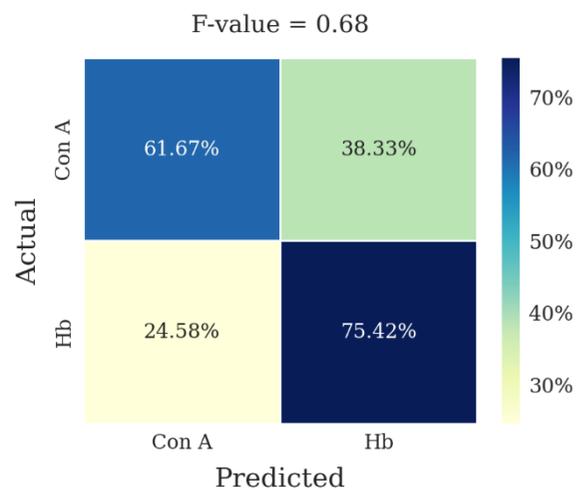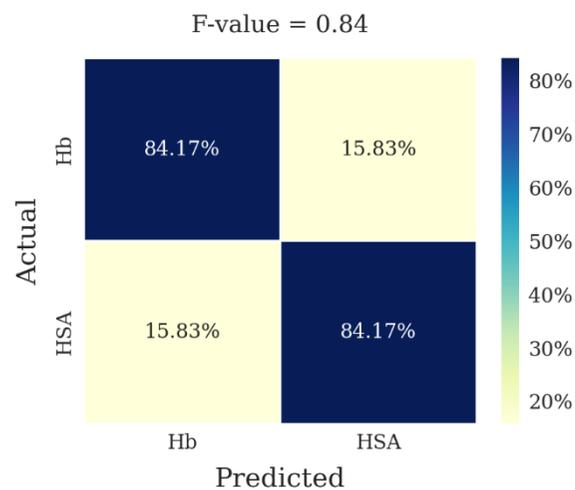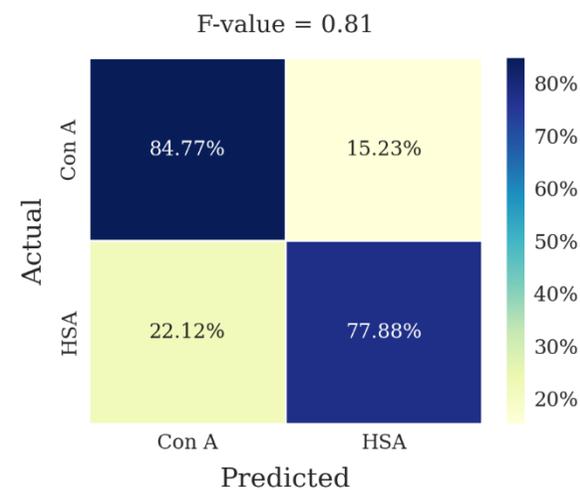

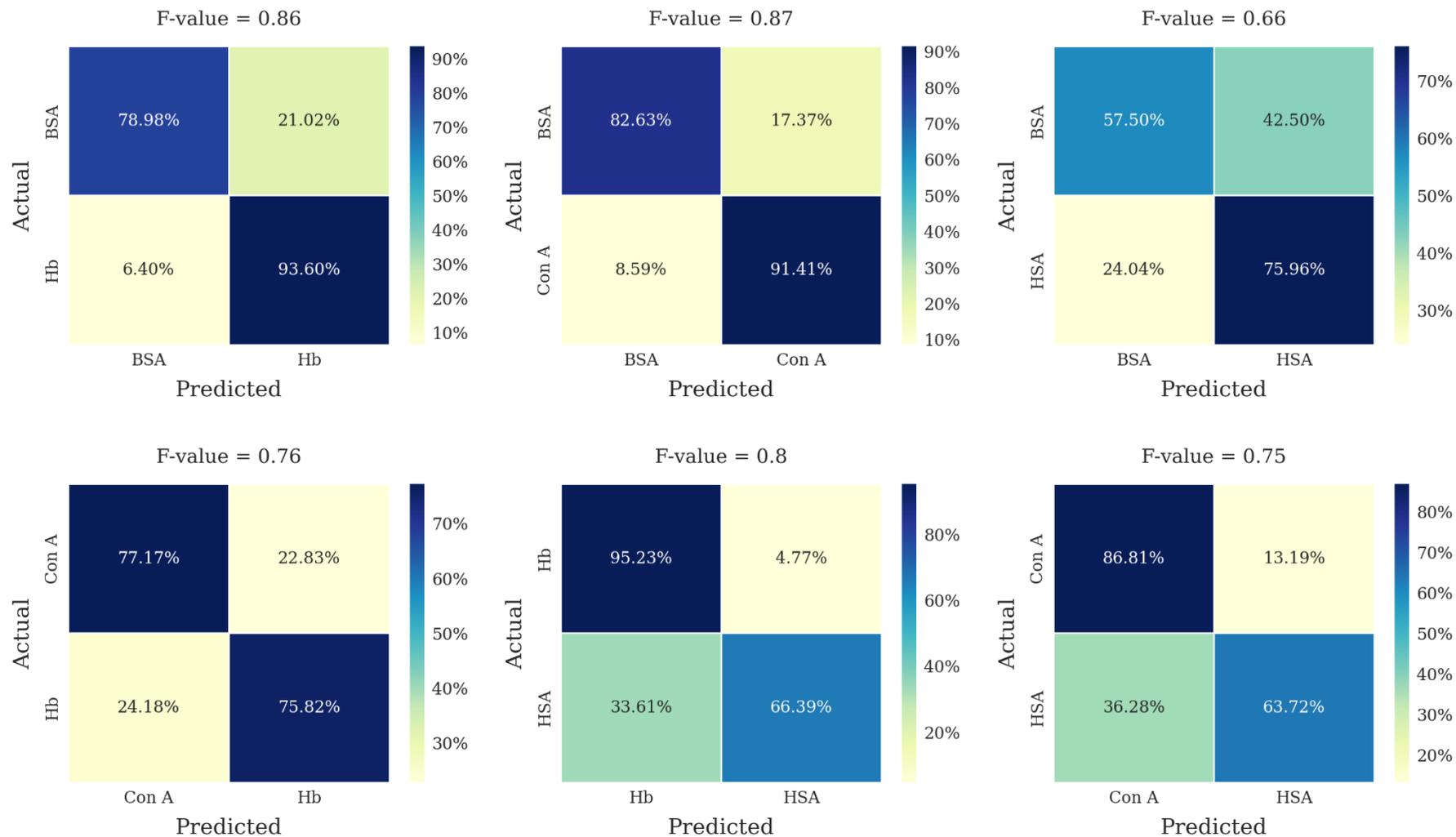

*Figure S13:* Confusion matrices of 2 types of proteins at 400mV using 200 ksps data (BW = 100 kHz) and 40 Msps data (BW = 10 MHz) employing Scheme 3. The numbers in the matrix element indicate the number of waveforms corresponding to that combination while the color of the element indicates the accuracy. The darker the color the higher the accuracy.

**Measurements done at 200 ksps (BW = 100 kHz)**

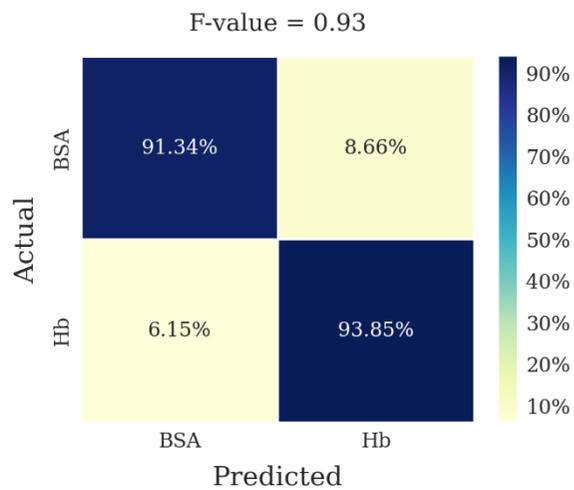
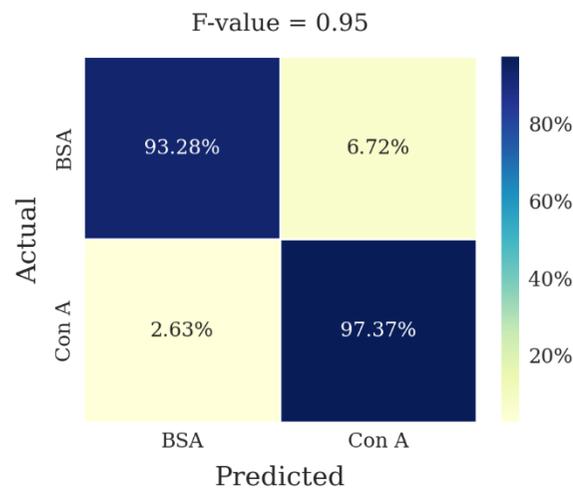
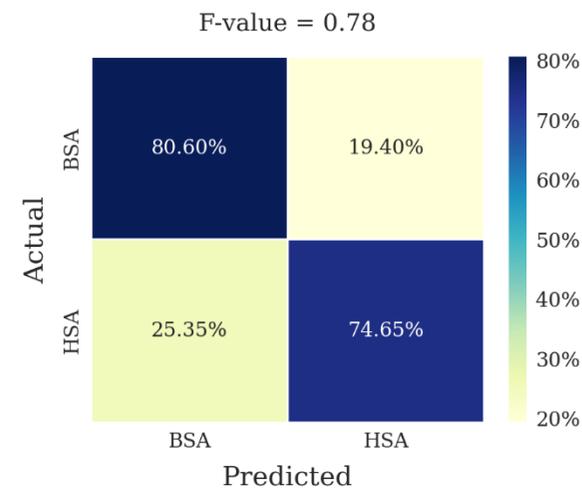
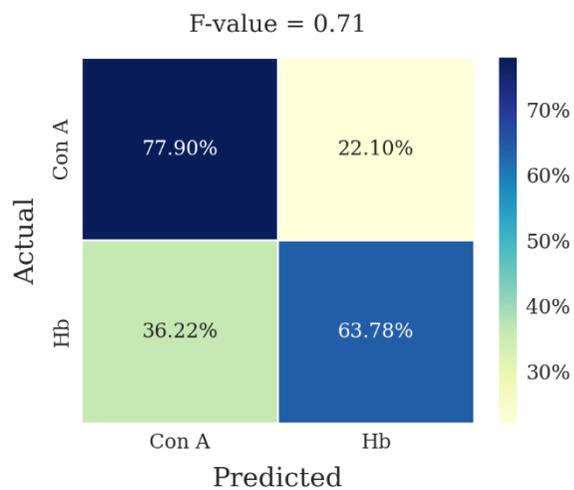
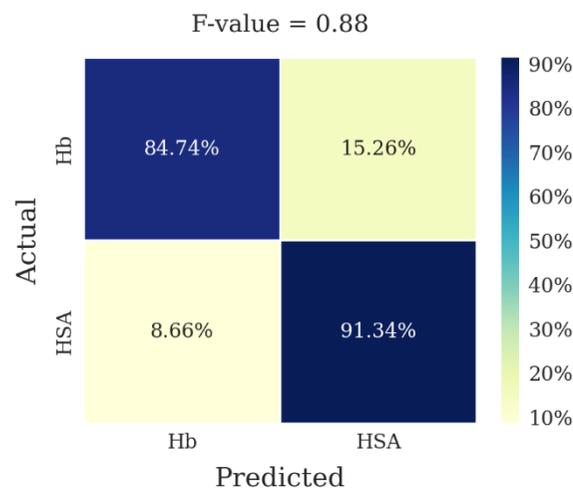
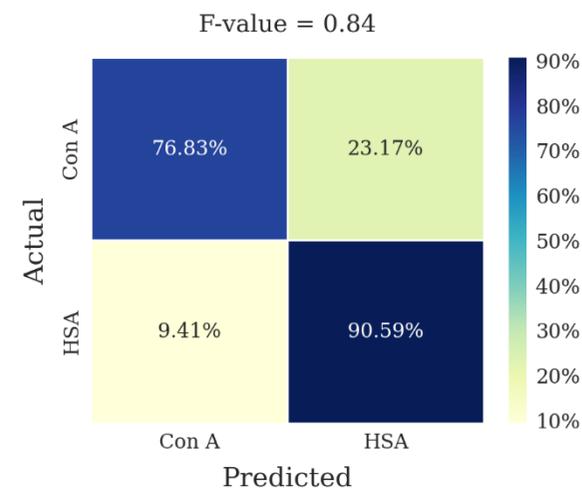

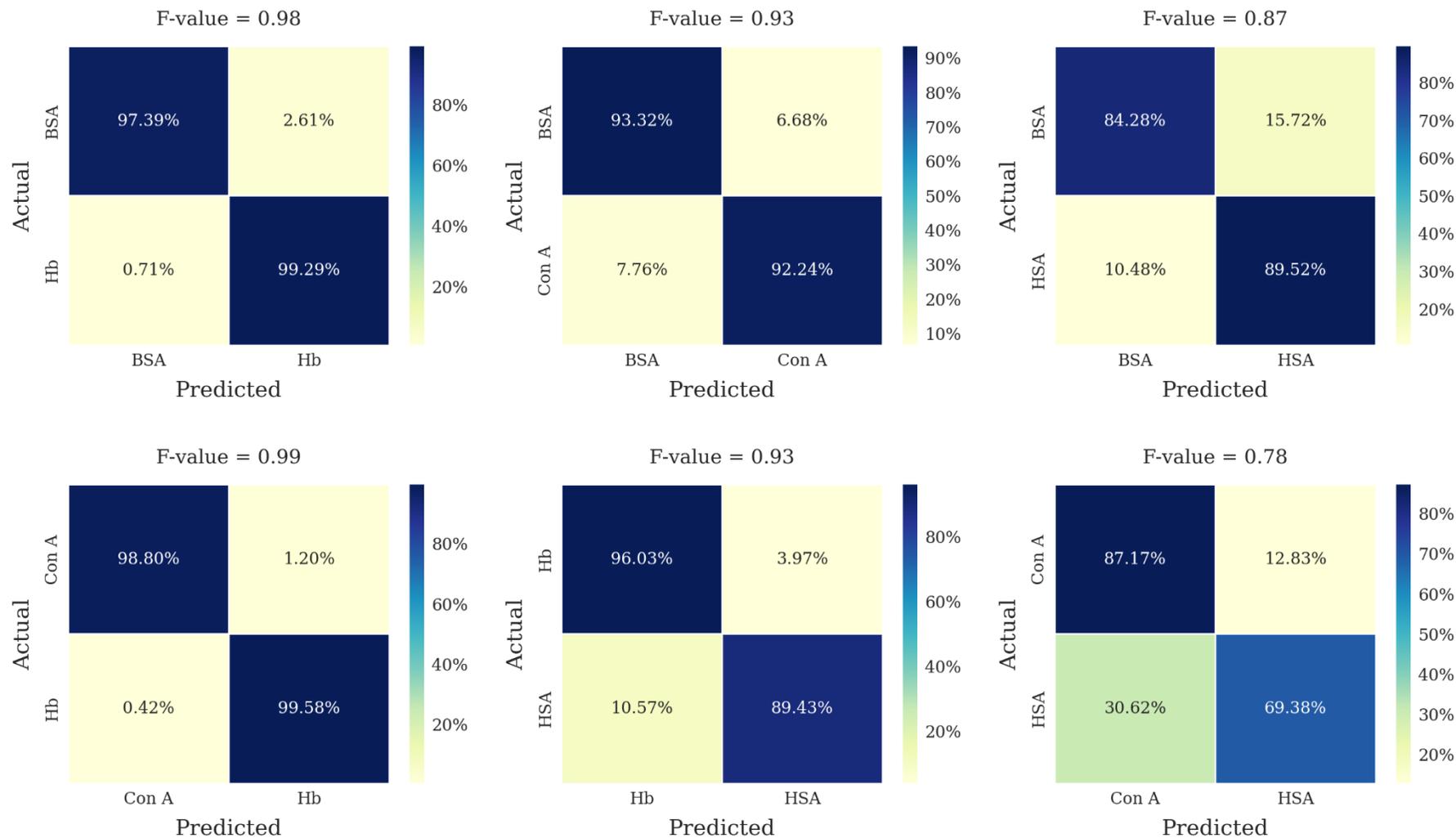

***Figure S14:*** *Confusion matrices of 2 types of proteins at 500mV using 200 ksps data (BW = 100 kHz) and 40 Msps data (BW = 10 MHz) employing Scheme 3. The numbers in the matrix element indicate the number of waveforms corresponding to that combination while the color of the element indicates the accuracy. The darker the color the higher the accuracy.*

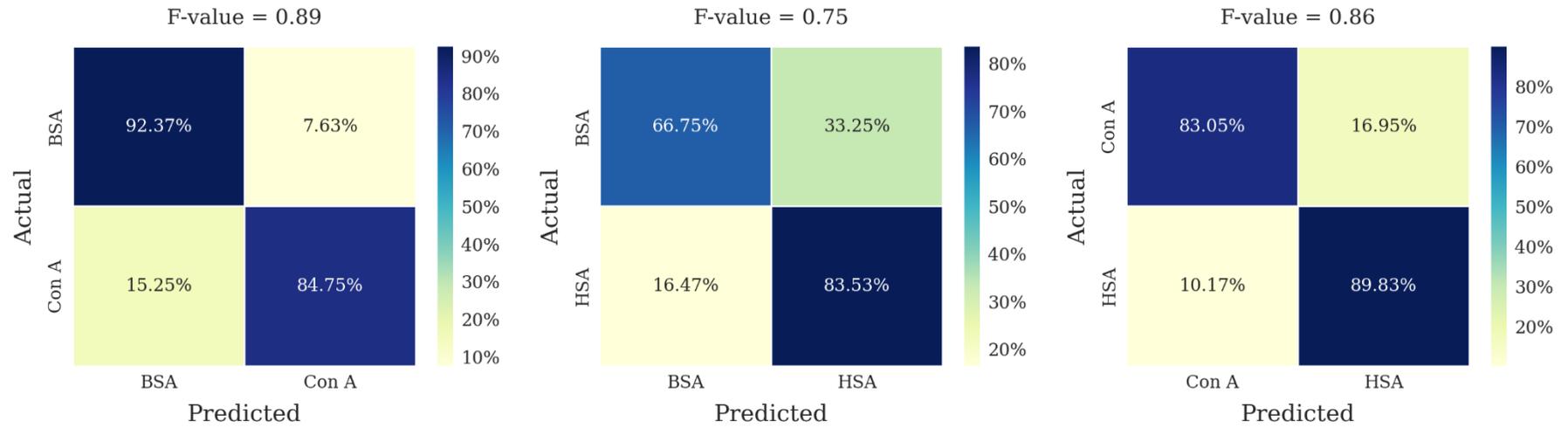

*Figure S15:* Confusion matrices of 2 types of proteins at 600 mV using 200 ksps data (BW = 100 kHz) employing Scheme 3. The numbers in the matrix element indicate the number of waveforms corresponding to that combination while the color of the element indicates the accuracy. The darker the color the higher the accuracy.

**Measurements done at 200 ksps (BW = 100 kHz)**

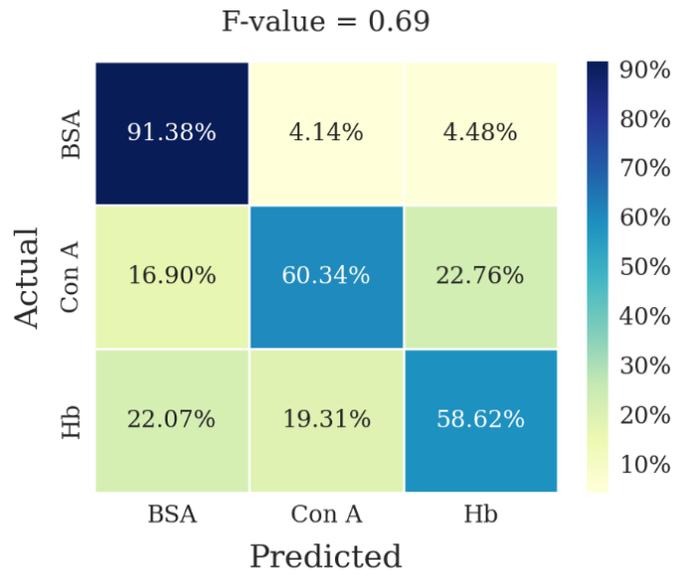
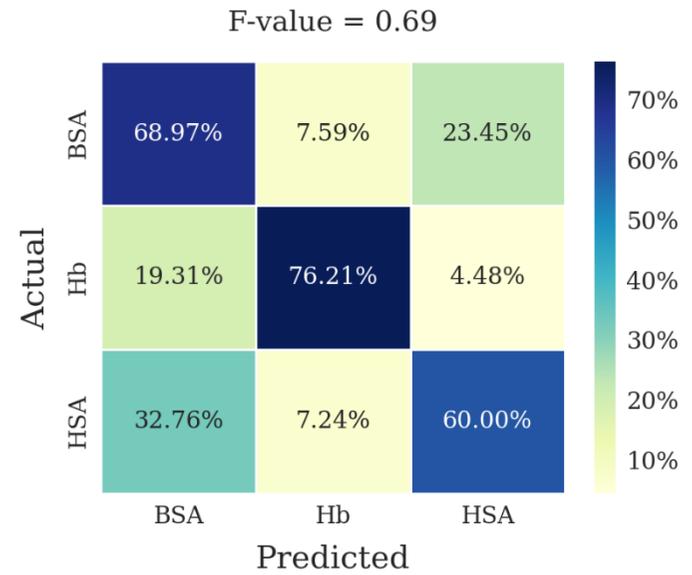
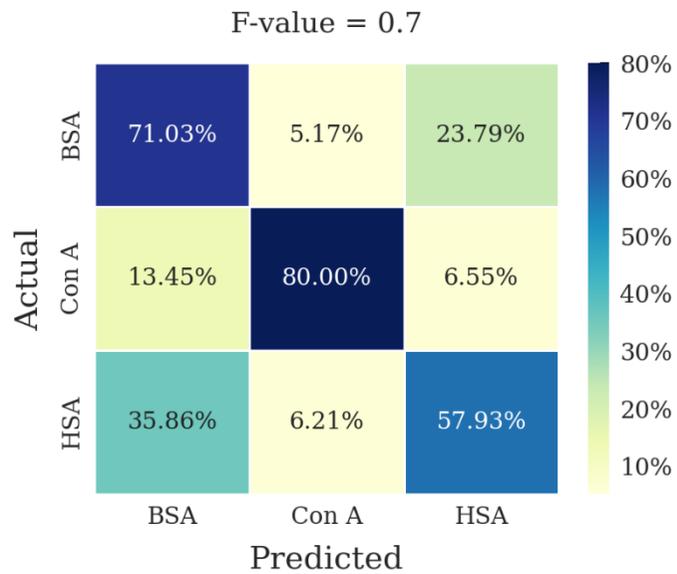
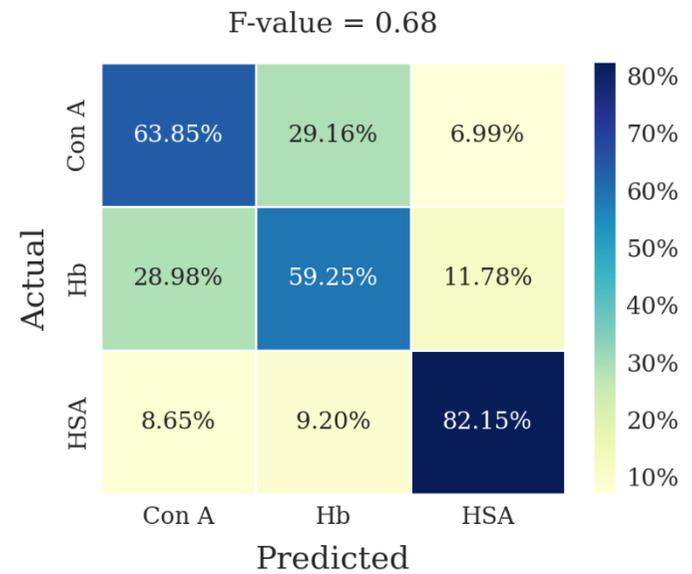

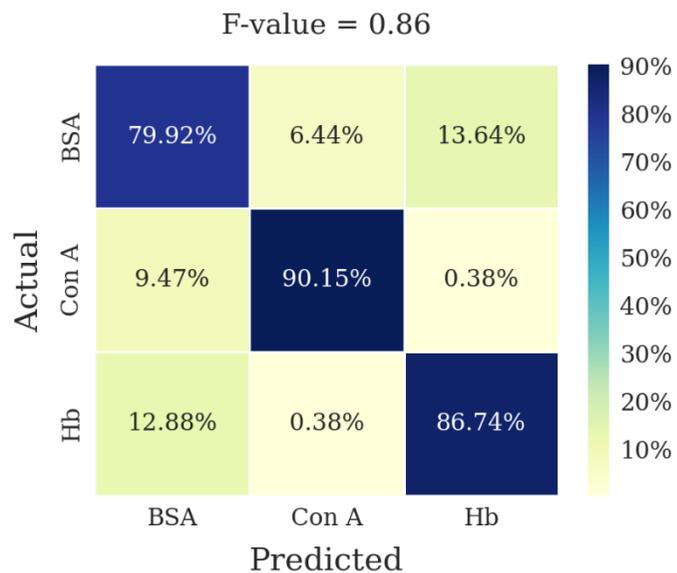
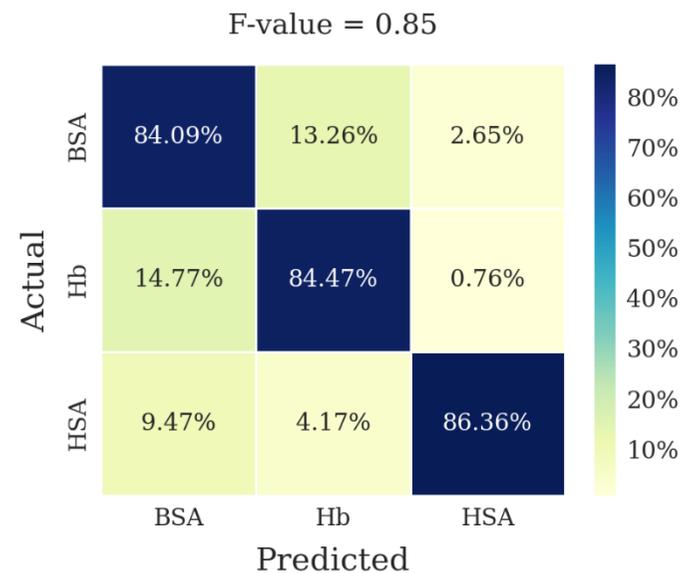
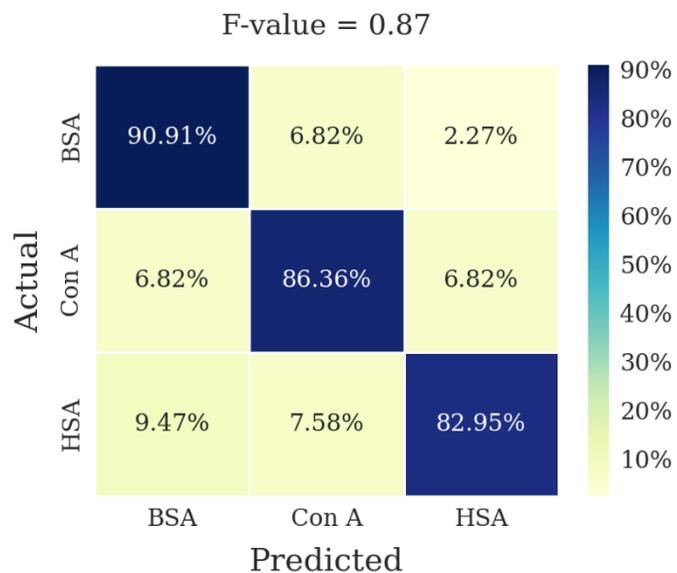
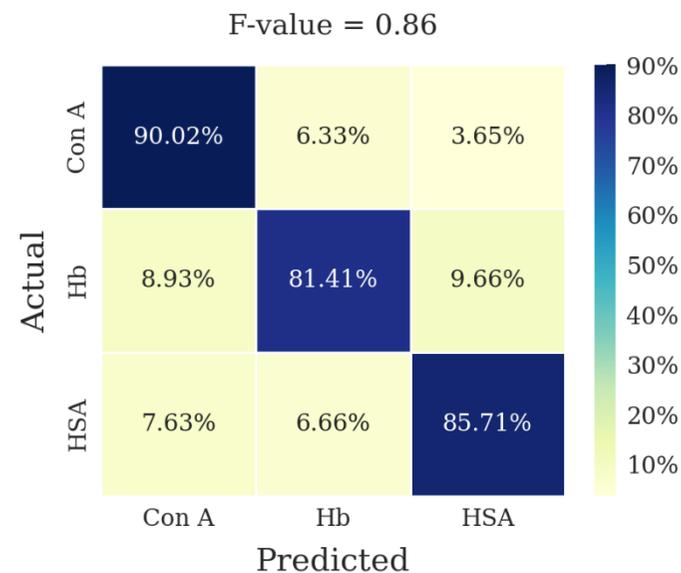

***Figure S16:*** *Confusion matrices of 3 types of proteins at 300mV using 200 ksps data (BW = 100 kHz) and 40 Msps data (BW = 10 MHz) employing Scheme 3. The numbers in the matrix element indicate the number of waveforms corresponding to that combination while the color of the element indicates the accuracy. The darker the color the higher the accuracy.*

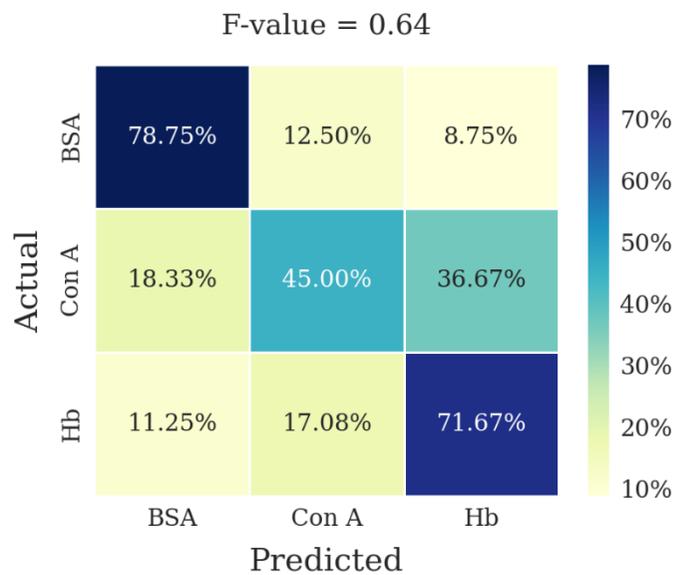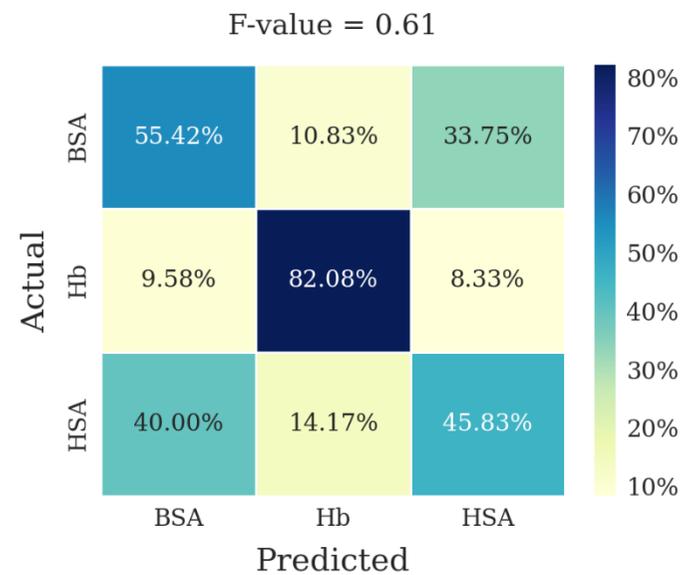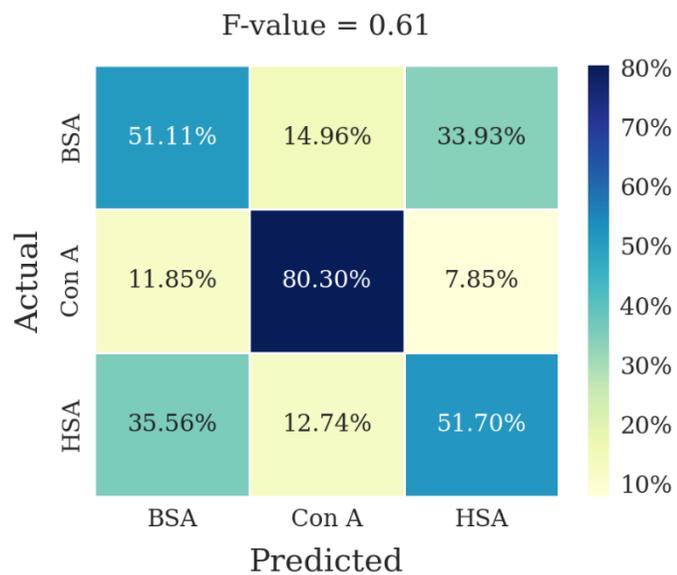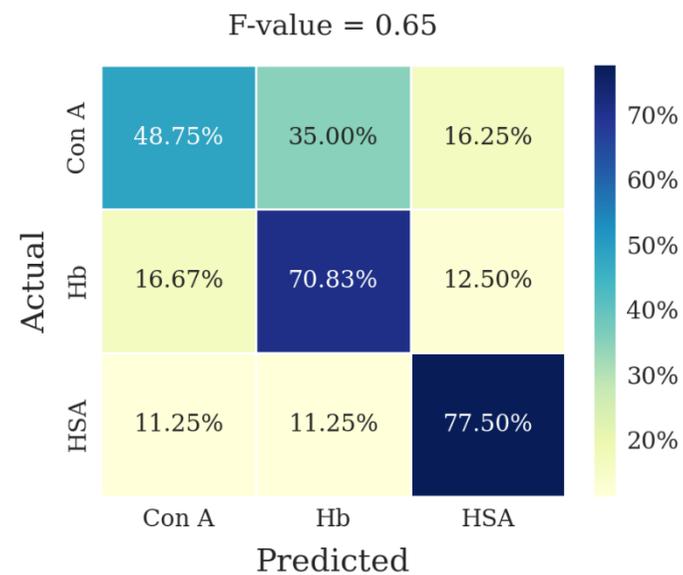

Measurements done at 200 ksps (BW = 100 kHz)

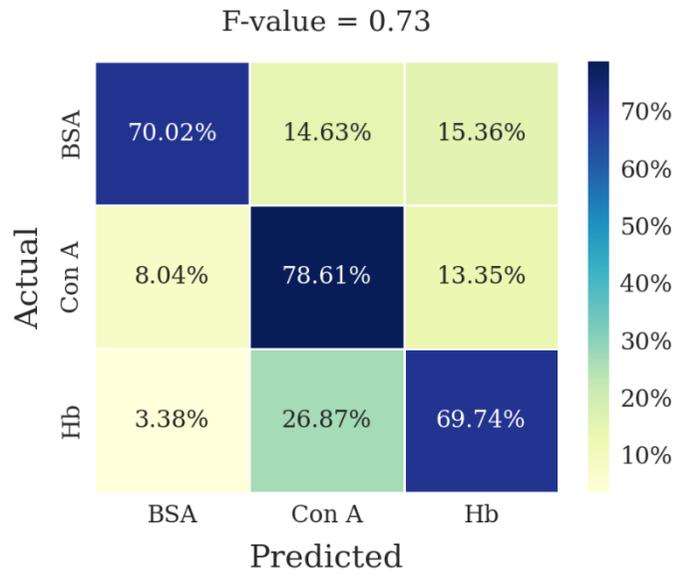
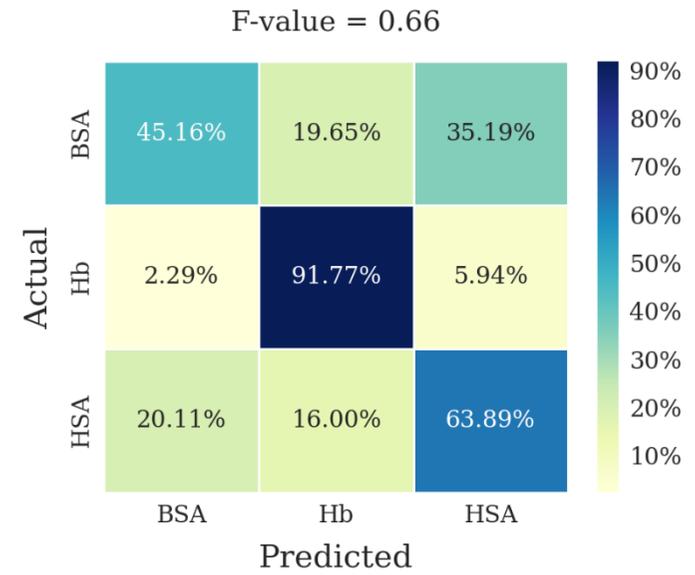
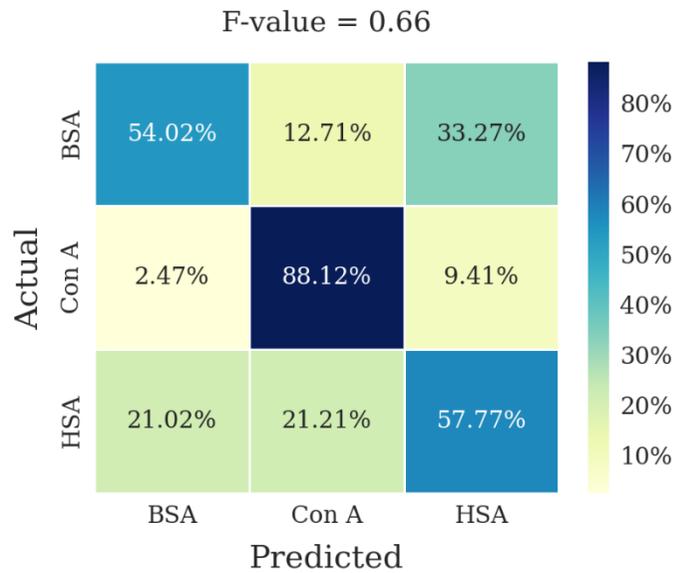
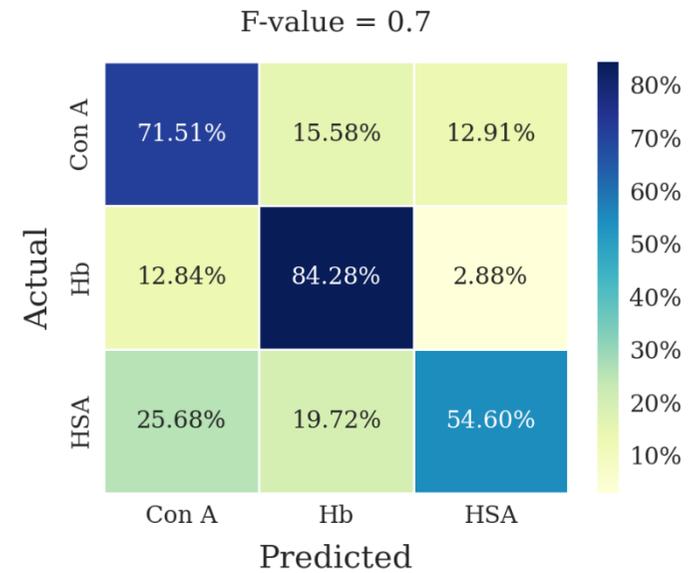

***Figure S17:*** *Confusion matrices of 3 types of proteins at 400mV using 200 ksps data (BW = 100 kHz) and 40 Msps data (BW = 10 MHz) employing Scheme 3. The numbers in the matrix element indicate the number of waveforms corresponding to that combination while the color of the element indicates the accuracy. The darker the color the higher the accuracy.*

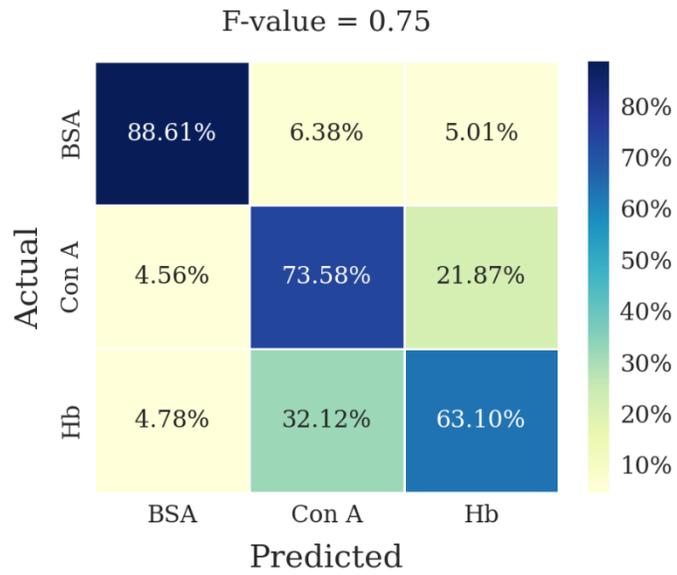
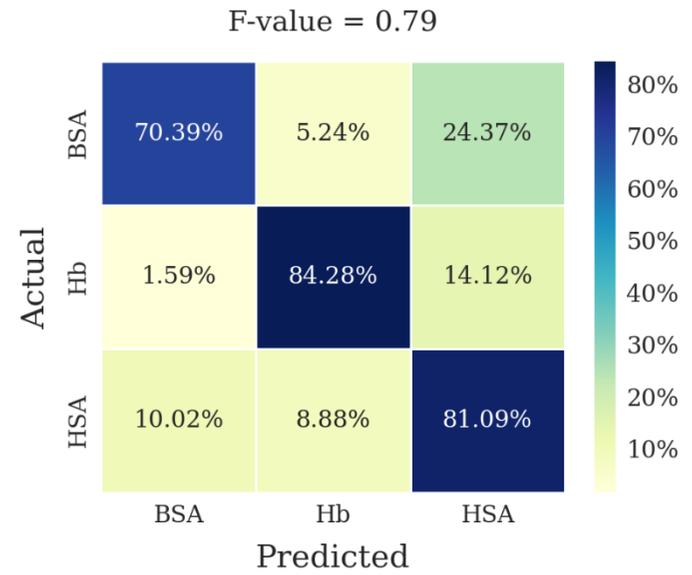
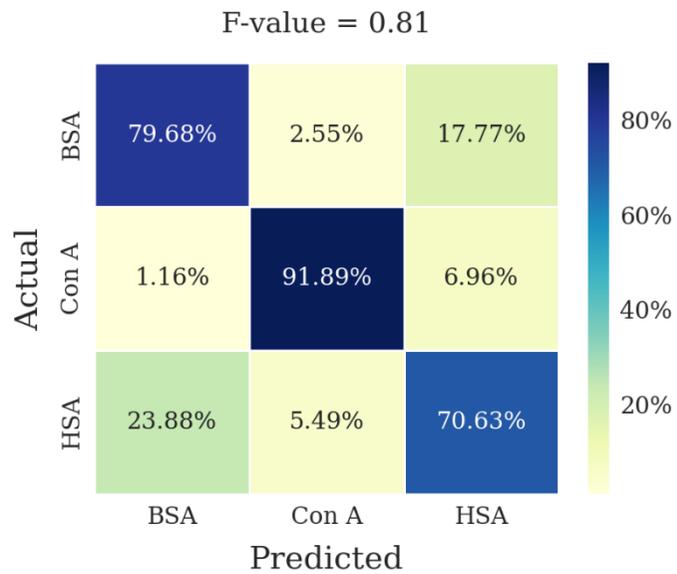
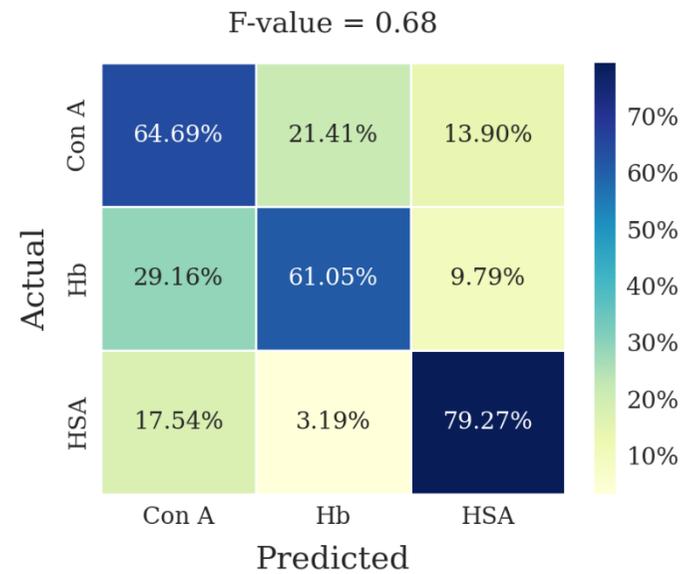

**Measurements done at 200 ksps (BW = 100 kHz)**

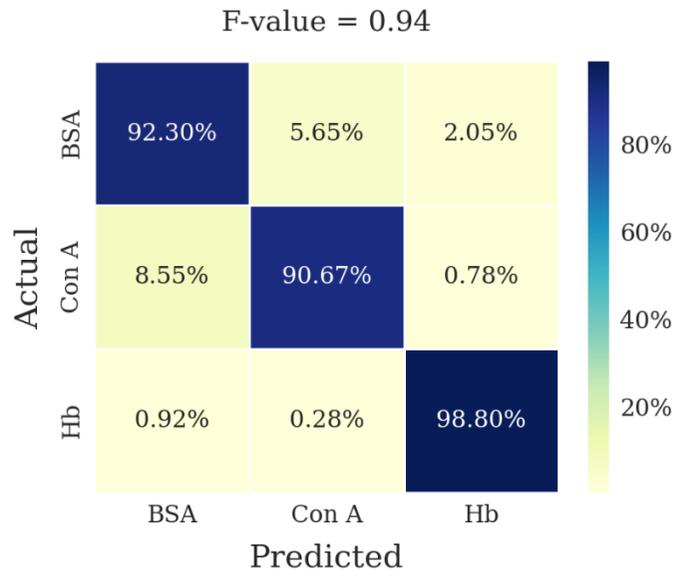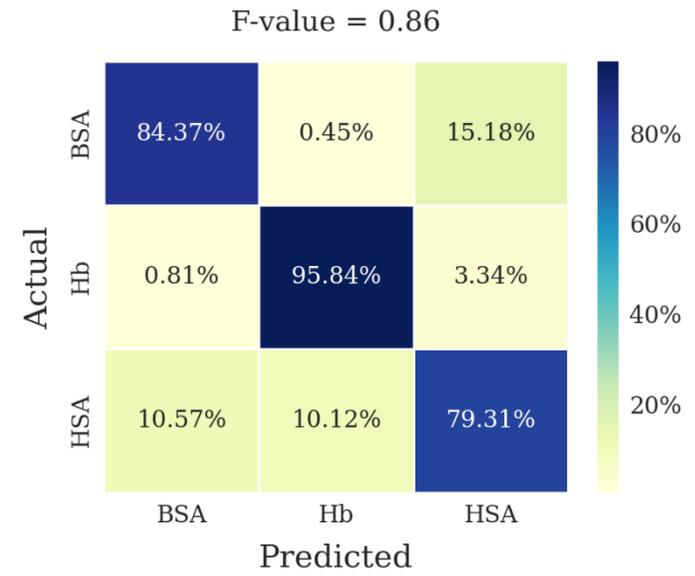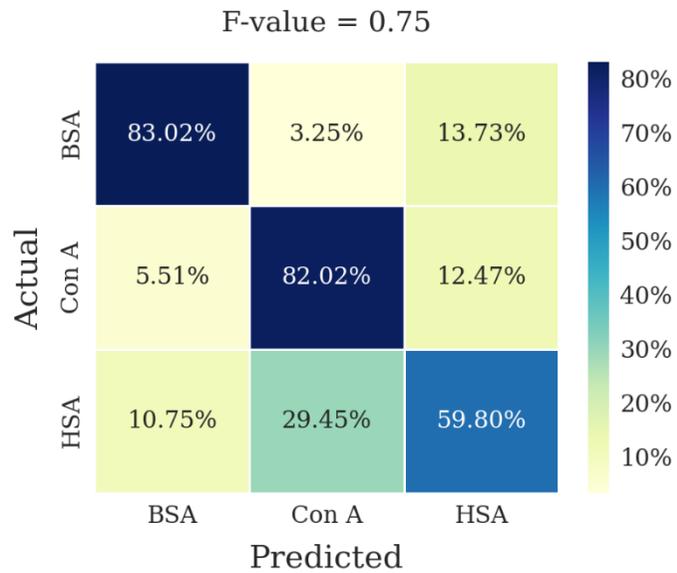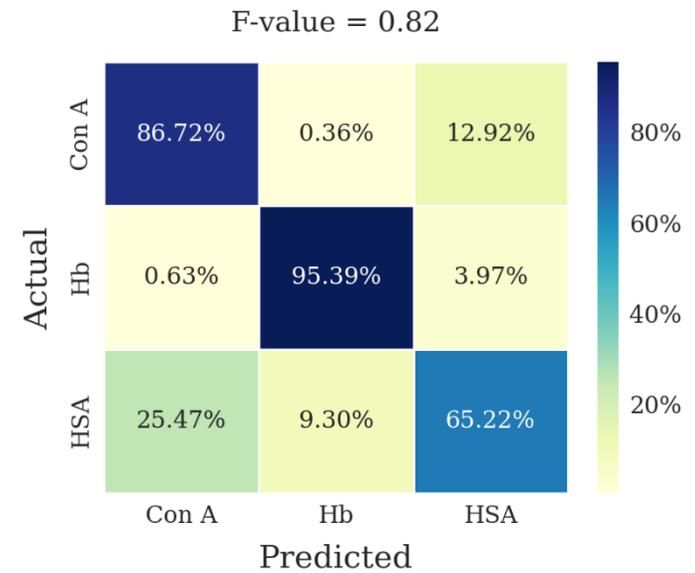

**Figure S18:** Confusion matrices of 3 types of proteins at 500mV using 200 ksps data (BW = 100 kHz) and 40 Msps data (BW = 10 MHz) employing Scheme 3. The numbers in the matrix element indicate the number of waveforms corresponding to that combination while the color of the element indicates the accuracy. The darker the color the higher the accuracy.

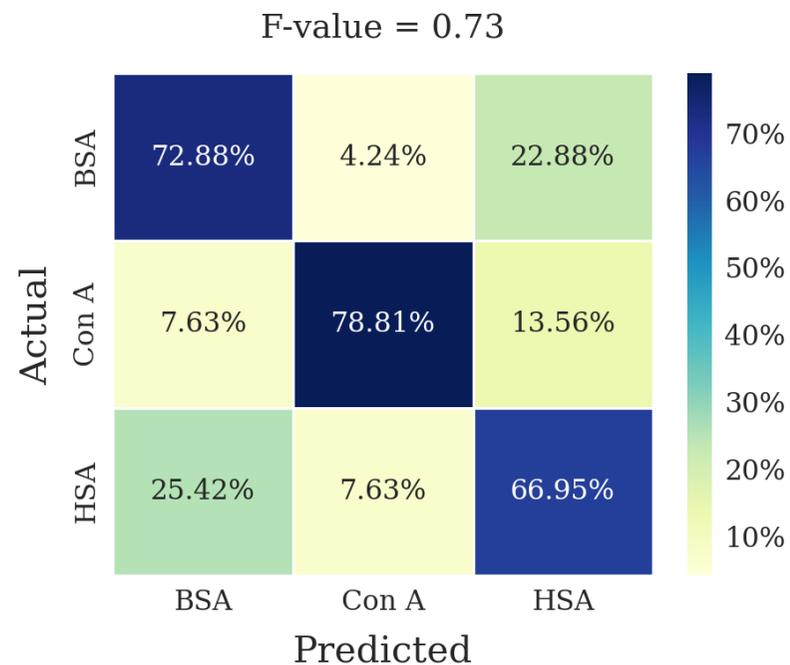

*Figure S19*: Confusion matrices of 3 types of proteins at 600mV using 200 ksps data (BW = 100 kHz) employing Scheme 3. The numbers in the matrix element indicate the number of waveforms corresponding to that combination while the color of the element indicates the accuracy. The darker the color the higher the accuracy.

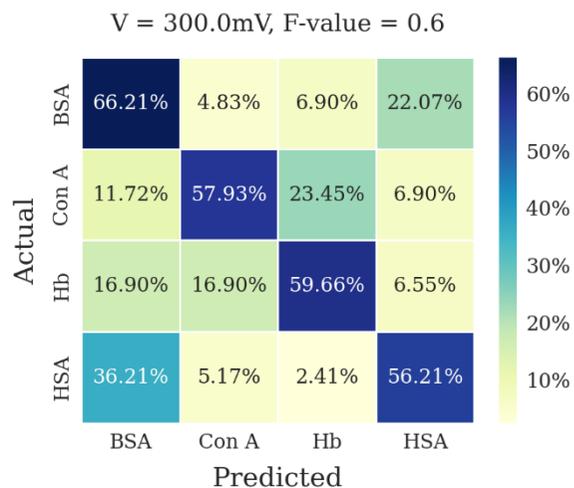 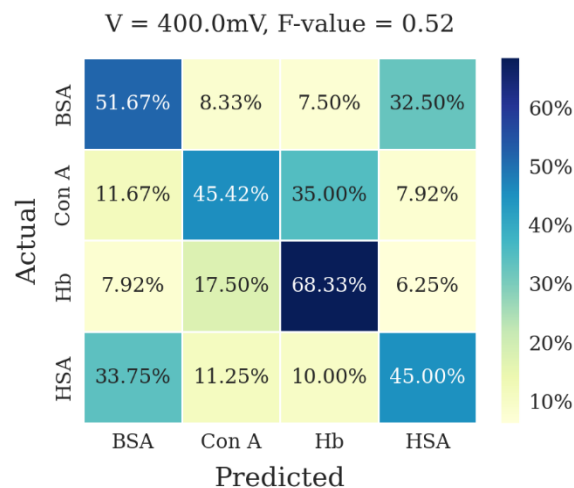 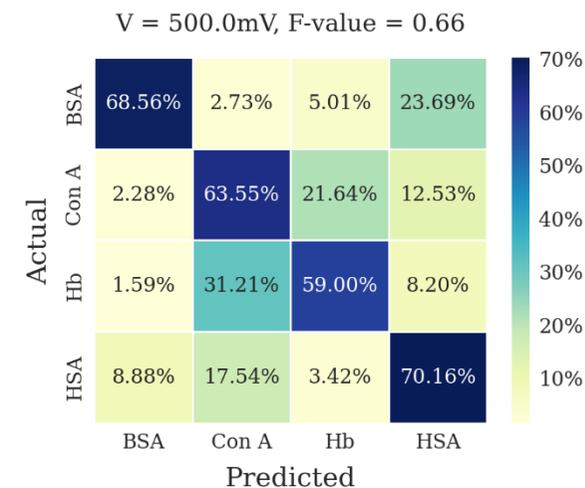
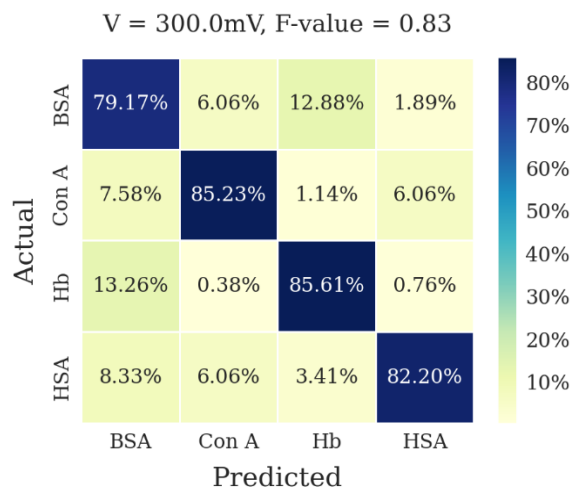 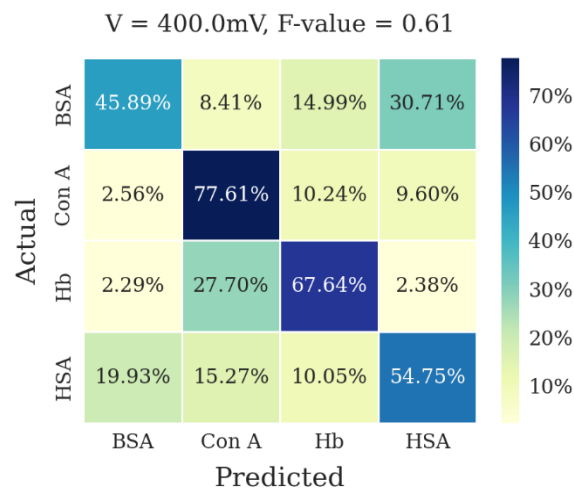 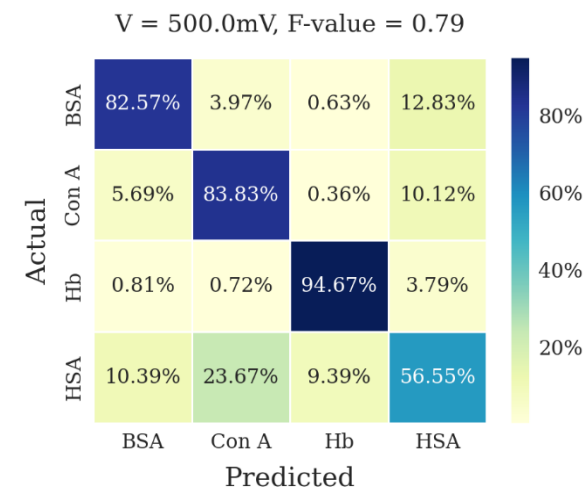

***Figure S20:*** *Confusion matrices of 4 types of proteins at 300mV, 400mV and 500mV using 200 ksps data (BW = 100 kHz) and 40 Msps data (BW = 10 MHz) employing Scheme 3. The numbers in the matrix element indicate the number of waveforms corresponding to that combination while the color of the element indicates the accuracy. The darker the color the higher the accuracy.*

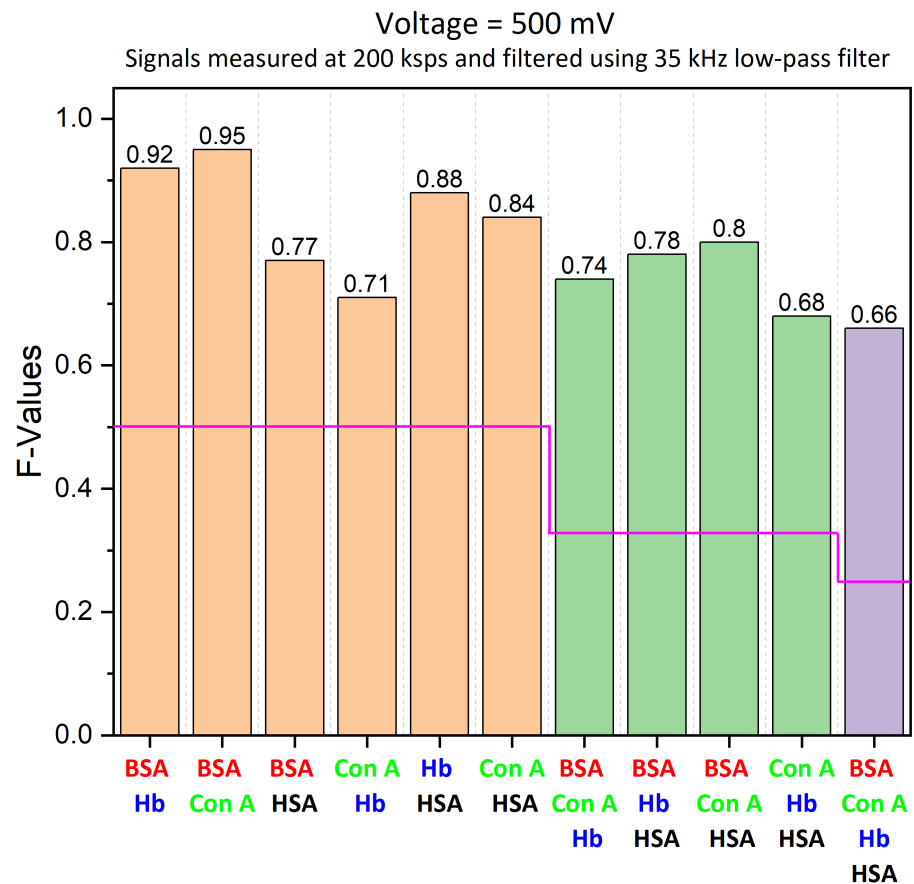 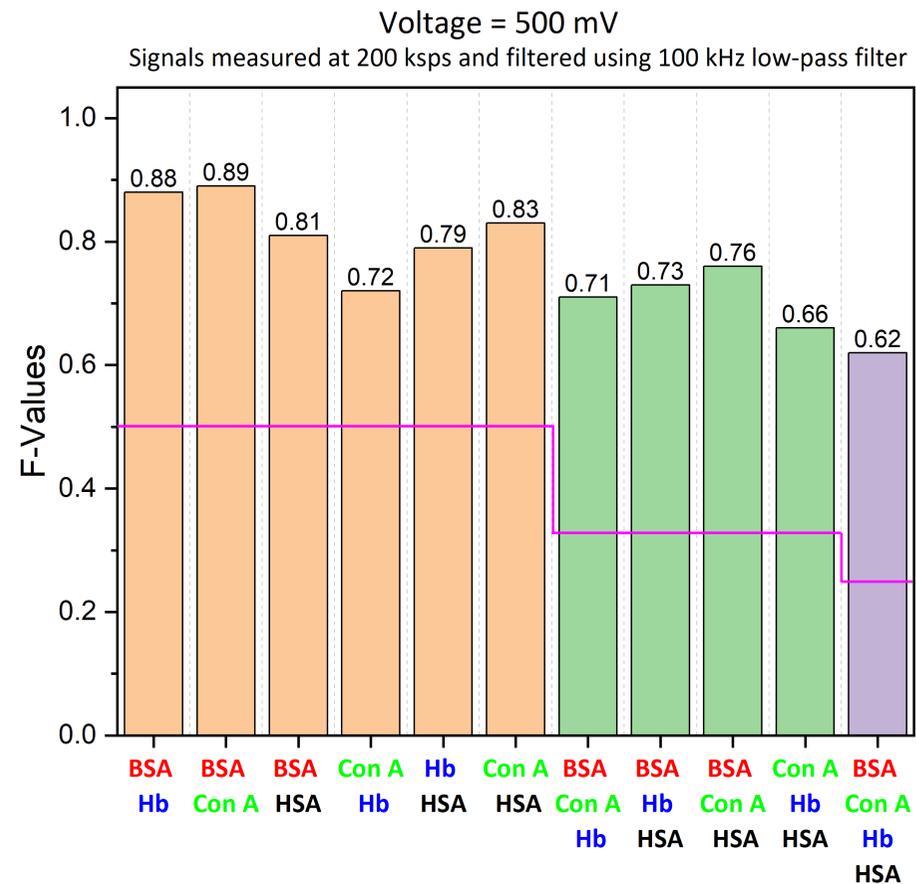

*Figure S21:* F-values in combinations of two, three and four types of proteins when data extraction was done after using a lowpass filter of (a) 35 kHz and (b) 100 kHz for data acquired at 200 ksps (BW = 100 kHz). The feature extraction was performed employing scheme 3.

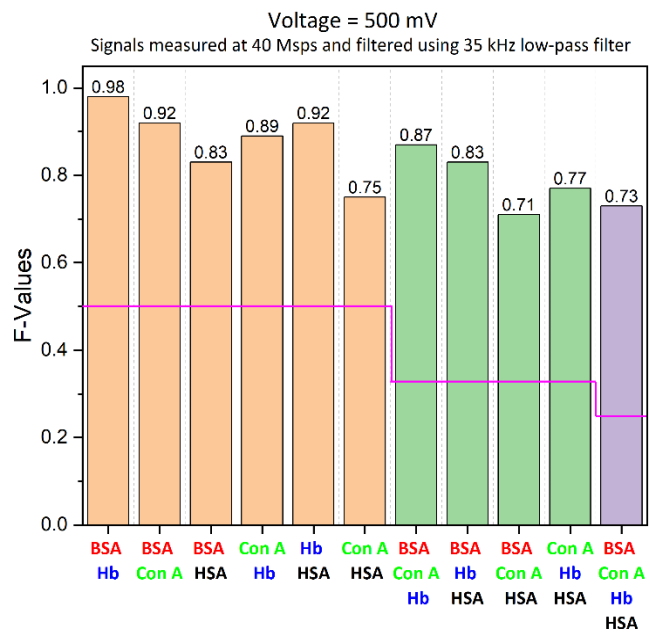 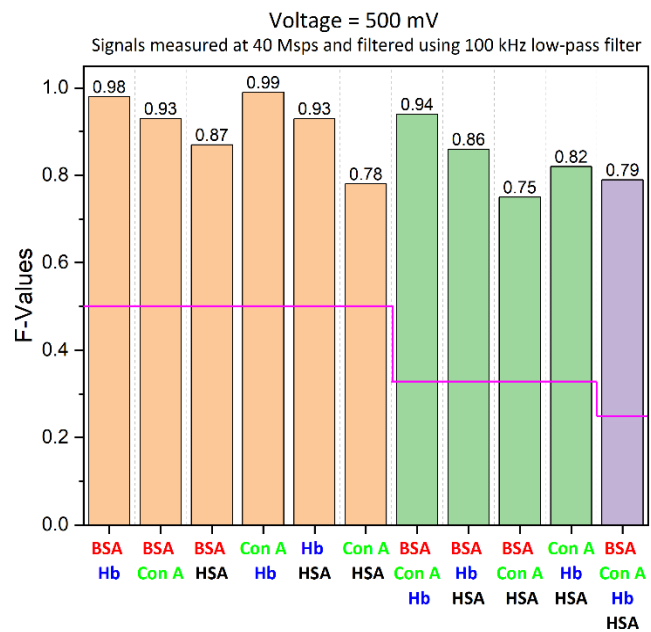 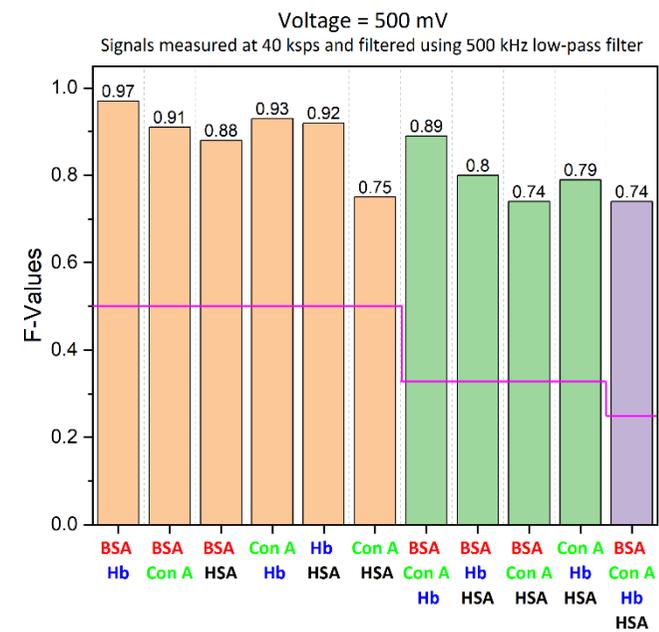

***Figure S22:*** *F- values in combinations of two, three and four types of proteins when data extraction was done after using a lowpass filter of (a) 35 kHz, (b) 100 kHz and (c) 500 kHz for data acquired at 40 Msps (BW = 10 MHz). The feature extraction was performed employing scheme 3.*